\documentclass[a4paper,11pt]{article}
\usepackage{jheppub} 
\usepackage{lineno}
\usepackage{graphicx}

\def\({\left(}
\def\){\right)}
\def\[{\left[}
\def\]{\right]}
\def\<{\langle}
\def\>{\rangle}

\def\nn {\nonumber}

\newcommand{\feynpdf}[2][80pt]{
    \vcenter{\hbox{\includegraphics[width=#1]{#2}}}}
\newcommand{\symfactor}[1]{\frac{1}{|\text{Aut}(#1)|}}

\title{Six Particles, Infinite Optimism: Towards Positivity Bounds for Six-Point Amplitudes}

\author[a]{Stefan Kremminger,}
\author[a]{and Andrew~J.~Tolley}

\affiliation[a]{Abdus Salam Centre for Theoretical Physics, Imperial College, London, SW7 2AZ, UK}

\emailAdd{stefan.kremminger24@imperial.ac.uk, a.tolley@imperial.ac.uk}

\abstract{The systematic generalisation of the well-established $S$-matrix positivity and bootstrap bounds from $2 \to 2$ scattering processes to higher-point amplitudes has, to date, remained elusive.
To address this, we consider a large class of tree-level UV-completions of a single-scalar EFT. Specifically, we consider the most general renormalisable $N$-scalar theory in $D=4$. By explicitly computing
tree-level amplitudes for four-, five- and six-point scattering and expanding at low energies, we can search for positivity relations among the expansion coefficients. We find that quite generally positivity bounds are unavailable due to contamination from a specific class of Feynman diagrams that are linear in sign-indefinite cubic couplings. However, by isolating a sector of the full crossing-symmetric low-energy amplitude that is unaffected by these sign-indefinite cubic couplings, we find that selected six-, five- and four-point Wilson coefficients organise into positive semi-definite matrices. While a violation of these positivity statements only implies that a given EFT is inconsistent with our class of UV completions, the result helps to guide towards what more general positivity bounds for higher-point interactions could, and could not, look like.}

\begin{document}
\maketitle
\flushbottom

\section{Introduction}

What interactions are consistent with the fundamental principles of nature? When working with an EFT, the naive answer is all local interactions consistent with the theory's symmetries. However, while a generic EFT includes all such interaction terms, it turns out that the majority of this parameter space is redundant: by assuming that the underlying physics fulfils basic properties like unitarity and causality, one can derive powerful constraints restricting this parameter space. These constraints fall broadly into $S$-matrix positivity bounds, which exploit the positive consequences of unitarity, and $S$-matrix bootstrap bounds, which impose the full nonlinear unitarity conditions. $S$-matrix positivity bounds were first developed in the late 1960s and early 1970s \cite{Yndurain:1969qm,Common:1970ck,Yndurain:1972ix}, but recent years have seen tremendous progress in these techniques, both theoretically and numerically.

Although this is now a well-developed field, with a multitude of approaches, most begin by using $S$-matrix analyticity to derive dispersion relations, which can then be further constrained by positivity and crossing symmetry
\cite{Pham:1985cr,Pennington:1994kc,Ananthanarayan:1994hf,Dita:1998mh,adams2006causality,deRham:2017avq,deRham:2017zjm,Remmen:2020vts,Zhang:2020jyn,Bellazzini:2020cot,Tolley:2020gtv,Caron-Huot:2020cmc,Sinha:2020win,Arkani-Hamed:2020blm,Li:2021lpe,Chiang:2021ziz,Alberte:2021dnj,deRham:2022hpx,deRham:2022sdl,Hong:2023zgm,DeAngelis:2023bmd,Wan:2024eto,Cheung:2025krg}.
The numerical $S$-matrix bootstrap programme further makes use of full unitarity to make stronger statements and is carried out via primal or dual methods, with primal methods constructing consistent amplitudes and dual methods deriving exclusion bounds
\cite{Paulos:2016fap,Paulos:2017fhb,Hebbar:2020ukp,Guerrieri:2021tak,Antunes:2023irg,He:2023lyy,Haring:2023zwu,Bhat:2024agd,Eckner:2024ggx,Guerrieri:2024jkn,
He:2025gws}.
Although the original primal bootstrap approach did not rely on dispersion relations, more recent approaches combine the two
\cite{deRham:2025vaq}.
Related bounds have also been derived for cross-section observables
\cite{Guerrieri:2024ckc,Correia:2025uvc}.
These programmes have been applied to particle physics through SMEFT and HEFT
\cite{Zhang:2020jyn,Chakraborty:2024ciu},
to quantum gravity, string theory, and the swampland
\cite{Guerrieri:2021ivu,Caron-Huot:2021rmr,Henriksson:2022oeu,Caron-Huot:2024lbf,Boisvert:2026sfh,Albert:2024yap},
to pions and large-$N$ QCD \cite{Albert:2022oes,Albert:2023jtd,Albert:2023seb,Albert:2026xyz}, and to parameterise the space of possible UV completions \cite{deRham:2026lvc}.

Two key ingredients in the derivation of these bounds are understanding the analytic structure of the $S$-matrix as a function of Mandelstam invariants and understanding the high-energy growth in the complex momentum plane. In the case of massive-particle scattering, the high-energy growth of amplitudes, or their Regge behaviour, is constrained by the Froissart--Martin and Jin--Martin bounds \cite{froissart1961asymptotic,jin1964connection}, which follow under the standard assumptions of analyticity, unitarity, a mass gap, and suitable polynomial boundedness. For scattering involving massless particles, particularly when gravitons are exchanged, additional assumptions are required (see, for example, \cite{Haring:2022cyf} for a discussion of Regge bounds with gravity). Knowing the high-energy growth, together with analyticity, allows one to close Cauchy contours in the complex energy plane to establish typically twice-subtracted dispersion relations \cite{Mandelstam1959analytic,ChewMandelstam1960}. 

When attempting to reproduce this success story for higher-point amplitudes, one immediately runs into problems. The number of independent Mandelstam invariants rapidly proliferates, making it very difficult to determine the analytic structure in Mandelstam variables. Unitarity is much more complicated, since multi-particle cuts involve products of amplitudes of different multiplicities, making it difficult to derive useful dispersion relations. Related to this, the Regge behaviour of higher-point amplitudes is less well studied, and results similar to the celebrated Froissart--Martin bound are not available.

Despite these technical challenges, there has been renewed interest in extending positivity bounds beyond four-particle scattering ($2 \rightarrow 2$) \cite{Chandrasekaran:2018qmx,Cheung:2025nhw,Cheung:2026lpv,Jeong:2026xzk}. While a general non-perturbative treatment based on dispersion relations still seems far away, some progress towards deriving new constraints on higher-point EFT interactions has been made. Recently, interesting positivity statements have been established \cite{Cheung:2025nhw,Cheung:2026lpv}, where moment inequalities are constructed from residues in specific scattering channels of different multiplicity. However, their direct translation into constraints on EFT expansion coefficients has so far been established in the planar limit.

To derive positivity bounds for non-planar amplitudes, namely amplitudes without a fixed planar ordering, we are therefore in a difficult position: the powerful tools used to derive four-particle bounds are unavailable, and a direct connection to residues is spoilt by channel mixing. To gain familiarity with the possible structures that arise in this context and guide us in the right direction, our approach in this paper is to study a specific class of renormalisable UV models, namely a general $N$-scalar theory in $D=4$. The downside of this approach is obvious: by imposing such a strong assumption on our UV model, a violation of any derived bound only implies inconsistency with this class of UV completions. However, having a situation in which we know everything allows us to better explore the kind of structures that seem to arise naturally, building intuition for what may hold in a more general setting.

In addition to constructing examples of explicit bounds that may or may not hold more generally, working within this well-defined context helps us to gauge expectations in a more general setting. As we outline in Section \ref{sect obstruction gabc}, we encounter sign-indefinite interactions that enter linearly in the six-point amplitude. Their presence significantly hinders the construction of six-point positivity bounds. Our attention therefore focuses on bounding the kinematic sector unaffected by these interactions. Within this kinematic region, a surprisingly natural structure emerges. Our lowest-order constraint connects the four-, five-, and six-point amplitudes in a single-scalar EFT:
\begin{equation}
    a^{(2)} c_D^{(4)}
    - \left(\frac{b^{(3)}}{4}\right)^2
    \geq 0 \,,
\end{equation}
for the EFT interactions $\frac{a^{(2)}}{8} (\partial \phi)^4$, $\frac{b^{(3)}}{4} \phi (\partial \phi)^2 (\partial\partial\phi)^2$, and $\frac{c_D^{(4)}}{2\,3!\,3!} \phi^3\Box^4\phi^3$. Using field redefinitions and integration by parts, we can equivalently express this as demanding positivity of the matrix 
\begin{equation}\label{eq intro result lowest order}
    \begin{pmatrix}
        \widetilde{a}^{(2)} & \widetilde{b}^{(3)}\\
        \widetilde{b}^{(3)}& c_D^{(4)}
    \end{pmatrix}\succeq 0 ,
\end{equation}
connecting the couplings in $\frac{\widetilde{a}^{(2)}}{8} \phi^2 \Box^2 \phi^2$, $\frac{\widetilde{b}^{(3)}}{2\,3!} \phi^3 \Box^3 \phi^2$, and $\frac{c_D^{(4)}}{2\,3!\,3!} \phi^3\Box^4\phi^3$. Similar statements also hold at higher order in the derivative expansion. In fact, quite remarkably, the Wilson coefficients of the four-, five- and six-point amplitude organise into the several positive semi-definite matrices. We find that \eqref{eq intro result lowest order} is a special case of positivity of the matrix of Wilson coefficients
\begin{equation}\label{result intro mixed orders}
\begin{pmatrix}
4a^{(2)}
&
b^{(3)}
&
b_A^{(4)}+\dfrac{3}{5}b_B^{(4)}
\\
b^{(3)}
&
4c_D^{(4)}
&
8c_E^{(5)}
\\
b_A^{(4)}+\dfrac{3}{5}b_B^{(4)}
&
8c_E^{(5)}
&
16c_J^{(6)}
+\dfrac{24}{5}c_H^{(6)}
+\dfrac{9}{25}c_I^{(6)}
\end{pmatrix}
\succeq 0.
\end{equation}
and all its principal minors. A second positive semi-definite matrix is 
\begin{equation}\label{result intro higher order}
        \begin{pmatrix}
            4 a^{(2)} & b_A^{(4)} &  b_B^{(4)} \\
            b_A^{(4)} & 16 c_J^{(6)} & 4 c_H^{(6)}\\
            b_B^{(4)} & 4 c_H^{(6)} & c_I^{(6)}
        \end{pmatrix} \succeq 0.
\end{equation}
The precise form of the EFT interactions associated with the above Wilson coefficients can be found in appendix \ref{appendix eft interactions}. 
The nonlinear bounds derived in this paper manifest as the positivity of the determinant or the principal minors of these two matrices with their size seemingly determined by the number of independent five-point interactions at each order. This suggests that they may be remnants of a more general consequence of unitarity. We stress that although superficially similar to the matrix positivity considered in \cite{Cheung:2025nhw,Cheung:2026lpv}, the bounds considered here 
apply to full amplitudes containing channel contributions
that are non-planar with respect to any fixed cyclic ordering. Thus, their positivity cannot be inferred from the underlying positivity of amplitude residues as considered in \cite{Cheung:2025nhw,Cheung:2026lpv}.

Taking inspiration from the analytic structure of four-point scattering, one would expect that statements analogous to \eqref{eq intro result lowest order} also connect Wilson coefficients with more derivatives. Although this seems to be the case at first glance, accessing the relevant terms at higher order in the derivative expansion seems problematic due to the presence of the Gram constraint, reflecting the fact that there can be at most $D$ linearly independent vectors in a $D$-dimensional vector space. This induces a polynomial relation among the generalised Mandelstam invariants for six-particle scattering. As we explain in the main text, both mixed-multiplicity and pure six-point candidate bounds arise at higher orders in the derivative expansion before imposing the Gram relation, but most do not survive once this is taken into account. Nonetheless, their appearance together with the survival of \eqref{result intro mixed orders} and \eqref{result intro higher order} provides an interesting stepping stone towards higher-point positivity bounds. The fact that this structure is not spoiled by channel-mixing is highly nontrivial and looks very promising for establishing analogous statements in a more general setting.

The rest of the paper is structured as follows: Section \ref{sect kinematic variables} discusses generalised Mandelstam variables for $N$-point scattering and how we circumvent the lack of generalised crossing symmetric variables. Section \ref{section model} introduces the model and identifies a first obstruction to bounding the model's Lagrangian. In Section \ref{section analyse amplitudes}, we then evaluate the four-, five-, and six-point amplitudes and derive bounds on six-point Wilson coefficients, including bounds that combine different multiplicities. We then apply these bounds to a single-scalar EFT in Section \ref{section eft}, before concluding in Section \ref{section conclusion}.

\section{Kinematic Variables}\label{sect kinematic variables}
\subsection{Defining Generalised Mandelstam Variables}
    Before discussing the specific model we work with, we must establish kinematic variables to represent the scattering amplitudes in. For four-point scattering, the canonical choice are the Mandelstam variables
    \begin{equation}\label{eq mandel def 4pt}
        s=-(p_1+p_2)^2 \qquad \qquad t=-(p_2+p_3)^2.
    \end{equation}
    For evaluating higher-point amplitudes, we define a set of generalised Mandelstam variables parameterising five- and six-particle kinematics. In line with the discussion given in \cite{Cheung:2025nhw}, we define 
    \begin{equation}\label{eq gen mandel def}
        s_i=-(p_1+...+p_{i+1})^2 \quad i=1,..,N-3 \qquad \qquad t_{ij}=-(p_i+...+p_j)^2 \quad 2\leq i< j \leq N-1.
    \end{equation}
    This yields five independent kinematic variables for five-point scattering, in line with the intuition that adding a new on-shell momentum to the scattering problem in $D=4$ should introduce three degrees of freedom. For six-point scattering there is an additional complication, as there cannot be five independent vectors in a four-dimensional space. This implies that of the nine variables resulting from \eqref{eq gen mandel def}, only eight are actually independent. The implicit relation follows from demanding that the Gram determinant has to vanish 
    \begin{equation}\label{eq gram constraint}
        \text{det} 
        \begin{pmatrix}
            0 & p_1 \cdot p_2 & p_1 \cdot p_3 & p_1 \cdot p_4 & p_1 \cdot p_5\\
            p_1 \cdot p_2 & 0 &p_2 \cdot p_3 & p_2 \cdot p_4  &p_2\cdot p_5\\
            p_1 \cdot p_3& p_2 \cdot p_3& 0 & p_3 \cdot p_4 & p_3 \cdot p_5\\
            p_1 \cdot p_4& p_2 \cdot p_4&p_3 \cdot p_4 &0&p_4 \cdot p_5\\
            p_1 \cdot p_5& p_2 \cdot p_5&p_3 \cdot p_5 &p_4 \cdot p_5 &0\\
        \end{pmatrix}=0.
    \end{equation}  
    The resulting equation is a quintic in Mandelstams but depends only quadratically on any individual variable. While one could in principle solve this constraint for any one variable, this introduces roots making calculations more cumbersome, and seems at odds with keeping the symmetries of the amplitude as manifest as possible. For this reason, we evaluate the six-point amplitude in terms of all nine variables, keeping \eqref{eq gram constraint} implicit. The obvious downside of this approach is that amplitude contributions that appear distinguishable due to their kinematic dependence may ultimately not be distinct. We will discuss the implications of this for our calculations in section \ref{sect imposing gram}.
  
\subsection{Lack of Crossing Symmetric Variables}
    In addition to Mandelstam variables \eqref{eq mandel def 4pt}, we may introduce 
    \begin{equation}\label{eq crossing sym vars 4point}
        x = s^2+t^2+u^2 \qquad \qquad y = s t u \, .
    \end{equation}
    Crossing symmetry is made manifest by expressing an amplitude in terms of these, as both $x$ and $y$ are invariant under permutations of the four momenta. Additionally, since the four-point amplitude depends polynomially on $x$ and $y$, one can determine the number of independent interactions in the EFT expansion at every order. For example, the only crossing symmetric contribution to four-point scattering at quartic order is $x^2$, while at sixth order two distinct combinations are possible.
    
    Similarly, it would be useful for our cause to express higher-point amplitudes in terms of fully crossing symmetric $N$-point variables. However, we are not aware of any similar variables in this context, which is why we resort to the generalised variables defined in \eqref{eq gen mandel def}. Crossing symmetry then implies that amplitudes depend only on fully crossing symmetric polynomials of these at every order. 
    
    To unambiguously read off the Wilson coefficient associated with any such expression, it is important that the set of polynomials is minimal. We therefore think of the k-th order derivative contribution as a vector space, with each independent polynomial as a basis element. This is similar to choosing a set of independent derivative interactions as the basis in which to describe an EFT. 
    The resulting polynomials are then linearly independent and, therefore, distinguishable from an EFT point of view. However, it is clear that there must be nonlinear algebraic relations between polynomials at different expansion orders, as their total number quickly surpasses the number of degrees of freedom in each amplitude. 
    
    An additional complication arises through the Gram constraint \eqref{eq gram constraint}, which captures linear relations between the external momenta in four dimensions. Imposing it may thereby induce linear relations between basis polynomials. To check whether a given set of polynomials remains independent, one has to check if they can be combined into something proportional to the Gram determinant. Imposing it can therefore only affect terms of quintic order or higher. For fifth-order polynomials one must test whether there is a combination of the chosen basis polynomials equal to the Gram determinant. If this is the case, the set is not independent after imposing the constraint. The analogue of this test at next order is to see if there is a weighted sum resulting to the determinant times some linear combination of Mandelstam variables. However, as both crossing symmetric combinations of six-point Mandelstam variables vanish identically at linear order by momentum conservation, one would not expect any correction from \eqref{eq gram constraint} for sixth-order contributions. We will comment on the implications of this for the specific model we are working with in section \ref{sect imposing gram}.
    
\section{EFT Matching and Obstructions to Positivity}\label{section model}
\subsection{UV Completion and EFT matching}
    To guide the search for positivity bounds beyond four-point scattering, we study a concrete UV theory. Consider the following Lagrangian involving $N$ scalar fields $\chi_a$
	\begin{equation}\label{lagrangian general renorm scalar}
		\mathcal{L}=\sum_a -\frac{1}{2} (\partial\chi_a)^2  -\sum_a \frac{m_a^2}{2} \chi_a^2 -\sum_{abc} \frac{g_{abc}}{3!} \chi_a \chi_b \chi_c - \sum_{abcd} \frac{\lambda_{abcd}}{4!} \chi_a \chi_b \chi_c \chi_d,
    \end{equation}
    which is perturbatively renormalisable in $D=4$. At tree-level, this theory may be regarded as a UV completion. Singling out one field $\phi$ and considering the regime $m_\phi\ll m_a$ where we neglect the light mass, we obtain
	\begin{equation}
    \begin{aligned} 
			\mathcal{L}=& -\frac{1}{2} (\partial \phi)^2 -\frac{g_\phi}{3!} \phi^3 -\frac{\lambda_\phi}{4!} \phi^4 - \sum_a \frac{1}{2} (\partial\chi_a)^2  -\sum_a \frac{m_a^2}{2} \chi_a^2\\
			&-\sum_{a} \frac{g_{a}}{2}\chi_a \phi^2 -\sum_a \frac{\lambda_a}{3!} \chi_a \phi^3 - \sum_{ab} \frac{g_{ab}}{2} \chi_a \chi_b \phi - \sum_{ab} \frac{\lambda_{ab}}{4} \chi_a \chi_b \phi^2\\
            &-\sum_{abc} \frac{g_{abc}}{3!} \chi_a \chi_b \chi_c -\sum_{abc} \frac{\lambda_{abc} }{3!}\chi_a \chi_b \chi_c \phi -\sum_{abcd}\frac{\lambda_{abcd}}{4!} \chi_a \chi_b \chi_c \chi_d.
    \end{aligned}
	\end{equation}
    The number of Latin indices on a given coupling constant thereby reflects how many heavy fields are involved in the interaction. By virtue of their definition the couplings with multiple indices may be taken as symmetric, e.g. $g_{ab}=g_{ba}$. We want to evaluate the scattering amplitudes of four, five and six $\phi$-particles at tree-level. To obtain the relevant action we solve the equation of motion for $\chi_a$ 
    \begin{equation}
        \Delta_a^{-1} \chi_a \equiv (\Box-m_a^2) \chi_a =  \frac{g_a}{2} \phi^2+\frac{\lambda_a}{3!} \phi^3 +\sum_b g_{ab} \chi_b \phi +\sum_b\frac{\lambda_{ab}}{2} \chi_b \phi^2 + \sum_{bc} \frac{g_{abc}}{2} \chi_b \chi_c +{\cal O}(\phi^5)
    \end{equation}
    iteratively and plug the result back into the action. Doing so, the induced interactions among external $\phi$-fields are captured by the tree-level effective action
	\begin{equation}\label{Lagrangian UV only phi}
    \begin{aligned}
	   \mathcal{L}=& -\frac{1}{2} (\partial \phi)^2-\frac{g_\phi}{3!} \phi^3 -\frac{\lambda_\phi}{4!} \phi^4 -\sum_a \frac{g_a^2}{8} \phi^2\Delta_a\phi^2\\
       &-\sum_a\frac{g_a \lambda_a}{2 \,3!} \phi^3 \Delta_a \phi^2 -\sum_{ab} \frac{g_a g_{ab} g_b}{8} \phi \(\Delta_a\phi^2\) \(\Delta_b \phi^2\)\\
       &-\sum_a \frac{\lambda_a^2}{2 \, 3!\, 3!} \phi^3\Delta_a \phi^3 -\sum_{ab} \frac{g_a \lambda_{ab} g_b}{16} \phi^2 \(\Delta_a\phi^2\)\(\Delta_b\phi^2\)\\
	   &- \sum_{ab} \frac{g_a g_{ab} \lambda_b}{2 \, 3!}\phi \(\Delta_a\phi^2\)\(\Delta_b\phi^3\)-\sum_{abc} \frac{g_a g_{ab} g_{bc} g_c}{8} \phi \(\Delta_a\phi^2\) \(\Delta_b \phi \(\Delta_c\phi^2\)\)\\
       &-\sum_{abc} \frac{g_a g_b g_c g_{abc}}{8\, 3!} \(\Delta_a\phi^2\) \(\Delta_b \phi^2\) \(\Delta_c \phi^2\)+{\cal O}(\phi^7).
    \end{aligned}
	\end{equation}

    \subsection{Obstructions to Positivity}\label{sect obstruction gabc}
    One straightforward obstruction to bounding the amplitude $\mathcal{A}_{6\phi}$ comes from the last term in \eqref{Lagrangian UV only phi}. The associated contribution is captured by Feynman diagrams of the form
    \begin{equation}
    \feynpdf[120pt]{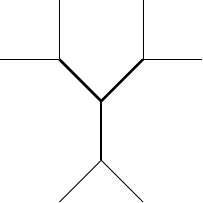} \propto g_{abc},
    \end{equation}
    for some pairing of external momenta. Summing over all diagrams of that topology and expanding for small energies, suppose that one specifies the first $n$ expansion coefficients in $\mathcal{A}_{6\phi}$ in order to bound higher-order contributions relative to these. As $g_{abc}$ enters linearly in the amplitude, the best one can do is use each piece of information to fix this coupling for a particular set of values $a,b,c$. Therefore, without setting an upper limit on the number of heavy fields in our model, no finite collection of data can be sufficient. As this coupling does not contribute to $\mathcal{A}_{4\phi}$ and $\mathcal{A}_{5\phi}$ at tree-level, lower-point amplitudes will not be helpful either in bounding its contribution.
    Therefore, in order to even have a chance of finding a statement, we restrict attention to terms in $\mathcal{A}_{6\phi}$ that do not get any contributions from $g_{abc}$. This strongly restricts the kinematic configurations that can be used to extract bounds for higher-point amplitudes and excludes, for example, situations where all but four external legs are soft.
\section{Expanding UV-Amplitudes}\label{section analyse amplitudes}
\subsection{Evaluating Contributions of Heavy Exchanges}\label{sect expand uv}
    Let us start by evaluating contributions to the amplitudes stemming from exchanges of heavy fields only. We will return to terms involving light propagators in section \ref{sect light exchanges uv}. \\
    For four-point scattering, we have
    \begin{equation}\label{eq expanded 4pt}
    \begin{aligned}
        \mathcal{A}_{4\phi}(s,t)=&\sum_{\sigma \in S_4} \symfactor{\Gamma}
         \feynpdf[120pt]{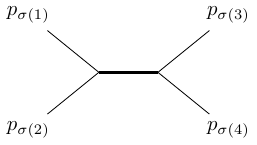} = \sum_a g_a^2\( \frac{1}{m_a^2-s} + \frac{1}{m_a^2-t} +\frac{1}{m_a^2+s+t} \)\\
         =& \sum_a\frac{3g_a^2}{m_a^2} + \sum_a \frac{2 g_a^2}{m_a^6}(s^2+st+t^2) - \sum_a\frac{3 g_a^2}{m_a^8}s t (s+t) + \sum_a \frac{2 g_a^2}{m_a^{10}}(s^2+st+t^2)^2 +{\cal O}(s,t)^5,
    \end{aligned}
    \end{equation}
    with heavy propagators denoted by a thick line. The symmetry factor $|\text{Aut}(\Gamma)|^{-1}$ assures that each inequivalent diagram with this topology contributes only once to the sum. Note the different summations in the above equation: $\sigma$ runs over permutations of the four external momenta while $a$ labels the propagating field. In line with the following discussion for five- and six-particle scattering, we introduce some notation for the crossing symmetric polynomials that arise in $\mathcal{A}_{4\phi}$. We define
    \begin{equation}
    \begin{aligned}
        k^{(n)} &= \frac{1}{2}\sum_{i<j} \( -(p_i+p_j)^2\)^n =(s^n+t^n+(-s-t)^n),
    \end{aligned}
    \end{equation}
    with $k^{(2)}$ and $k^{(3)}$ proportional to the crossing symmetric variables $x=s^2+t^2+u^2$ and $y=s t u$ typically used for four-point scattering. The factor $1/2$ is added to avoid the double counting otherwise implied by momentum conservation. The Wilson coefficient multiplying $k^{(n)}$ is denoted as $\alpha^{(n)}$. We therefore have
    \begin{equation}
        \mathcal{A}_{4\phi}= \alpha^{(0)} + \alpha^{(2)} k^{(2)} +\alpha^{(3)} k^{(3)} + \alpha^{(4)} k^{(4)}+ {\cal O}(s,t)^5,
    \end{equation}
    with four-point expansion coefficients
    \begin{equation}
        \alpha^{(0)}= \sum_a \frac{3 g_a^2}{m_a^2} \qquad \alpha^{(2)}=\sum_a \frac{g_a^2}{m_a^6} \qquad \alpha^{(3)}=\sum_a \frac{g_a^2}{m_a^8} \qquad \alpha^{(4)}=\sum_a \frac{g_a^2}{m_a^{10}}.
    \end{equation}
   We stress here that, even though all different $k^{(n)}$ can be expressed in terms of the crossing symmetric variables $x$ and $y$, they nonetheless give distinct contributions to $\mathcal{A}_{4\phi}$. This will yield the crucial link into the EFT regime, where contributions proportional to  $k^{(n)}$ can equally be isolated unambiguously.
   
   Notice the manifest positivity of the Wilson coefficients $\alpha^{(n)}$. Given we fix $\alpha^{(2)}$, all higher-order terms in the expansion will be bounded relative to it. Introducing the energy scale $\Lambda \leq m_a$, we see that
    \begin{equation}
        \alpha^{(n)}= \sum_a \frac{g_a^2}{m_a^{2n+2}}\leq \frac{1}{\Lambda^{2n-4}} \sum_a \frac{g_a^2}{m_a^6} =\frac{\alpha^{(2)}}{\Lambda^{2n-4}}.
    \end{equation}
    This captures how the regular story of four-point positivity manifests in the particular class of UV-complete theories considered here. The full scattering amplitude receives contact terms which prevent a direct bound on $\alpha^{(0)}$. The higher order terms admit a nonlinear bound which can be stated as positive semi-definiteness of the matrix
    \begin{equation}\label{eq bound 4point nonlin}
\begin{pmatrix}
\alpha^{(2)} & \alpha^{(3)}
\\
\alpha^{(3)} & \alpha^{(4)}
\end{pmatrix}
\succeq 0 \, .
\end{equation}
We stress that this is specific to this class of UV completions, with the more general bounds given in \cite{Tolley:2020gtv,Caron-Huot:2020cmc}.
    
For five-particle scattering we must consider two types of heavy-exchange diagrams,
    \begin{equation}
        \feynpdf[120pt]{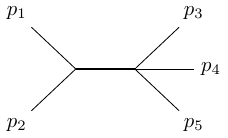} = \sum_a g_a \lambda_a \frac{1}{m_a^2-s_1}
\end{equation}
\begin{equation}
         \feynpdf[120pt]{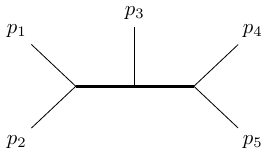}=-\sum_{ab} \frac{g_a g_{ab} g_b}{2} \frac{1}{m_a^2-s_1} \frac{1}{m_b^2-s_2}
         \end{equation}
as well as permutations of their external momenta. We again expand the five-point amplitude and express the result in terms of some crossing symmetric polynomials.
\begin{eqnarray}\label{eq 5pt in pictures uv}
            \mathcal{A}_{5\phi}=&&\sum_{\sigma \in S_5} \symfactor{\Gamma_1}
            \feynpdf[120pt]{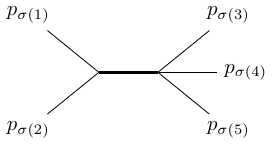}+ \symfactor{\Gamma_2}
        \feynpdf[140pt]{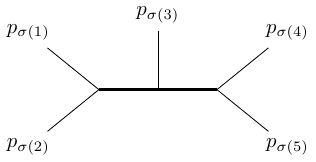} \nonumber \\
        =&& \beta^{(0)} + \beta^{(2)} q^{(2)}+ \beta^{(3)}  q^{(3)} + \beta_A^{(4)}  q_A^{(4)}+ \beta_B^{(4)} q_B^{(4)} + {\cal O}(s_i,t_{ij})^5.
        \end{eqnarray}
It turns out that these expansion coefficients can be written most naturally in terms of the shifted coupling 
\begin{equation}\label{eq def shifted lambda}
        \widetilde{\lambda}_a\equiv\lambda_a -\sum_b \frac{g_{ab}g_b}{m_b^4} (m_a^2+3m_b^2).
\end{equation}
Precise expressions for the Wilson coefficients and crossing symmetric
polynomials are
\begin{eqnarray}
\beta^{(0)}
&=&
10 \sum_a \frac{g_a \lambda_a}{m_a^2}
-15\sum_{ab} \frac{g_a g_{ab} g_b}{m_a^2 m_b^2}
\nonumber\\
\beta^{(2)}
&=&
\sum_a \frac{g_a \lambda_a}{m_a^6}
-\frac{1}{2}\sum_{ab}
\frac{g_a g_{ab}g_b}{m_a^6 m_b^6}
m_a^2(6m_a^2+m_b^2)
\nonumber\\
\beta^{(3)}
&=&
\sum_a\frac{g_a \widetilde{\lambda}_a}{m_a^8}
\,,\qquad
\beta_A^{(4)}
=
\sum_a \frac{g_a \widetilde{\lambda}_a}{m_a^{10}}
\nonumber\\
\beta_B^{(4)}
&=&
-\sum_{ab} \frac{g_a g_{ab}g_b}{m_a^6 m_b^6}
\nonumber\\[3pt]
q^{(2)}
&=&
\sum_{i<j} \(-(p_i+p_j)^2\)^2
\,,\qquad
q^{(3)}
=
\sum_{i<j} \(-(p_i+p_j)^2\)^3
\nonumber\\
q_A^{(4)}
&=&
\sum_{i<j} \(-(p_i+p_j)^2\)^4
\nonumber\\
q_B^{(4)}
&=&
\frac{1}{2}\sum_{\substack{i<j\\ k\neq i,j}}
\(-(p_i+p_j)^2\)^2
\(-(p_i+p_j+p_k)^2\)^2 .
\label{eq coefficients and polynomials 5point}
\end{eqnarray}
Here, the quartic coupling $\widetilde{\lambda}_a$ is defined in
\eqref{eq def shifted lambda}.
    We see that at fourth order in the derivative expansion there are two distinguishable contributions to the amplitude, captured by the two independent polynomials $q_A^{(4)}$ and $q_B^{(4)}$. We add a factor of $1/2$ to the definition of $q_B^{(4)}$ as any pairing of momenta contributes twice by momentum conservation.
 
    Finally, we repeat this procedure for six particles, with the amplitude given in terms of the following Feynman diagrams and permutations of their external legs.
\begin{equation}
        \feynpdf[120pt]{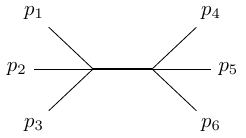}= \sum_a \frac{\lambda_a^2}{2} \frac{1}{m_a^2-s_2} 
\end{equation}
\begin{equation}
        \feynpdf[120pt]{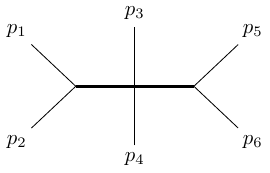}=-\sum_{ab} \frac{g_a \lambda_{ab} g_b}{2} \frac{1}{m_a^2-s_1} \frac{1}{m_b^2-s_3}
\end{equation}
\begin{equation}
        \feynpdf[120pt]{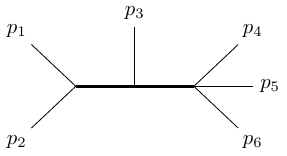}=-\sum_{ab} g_a g_{ab} \lambda_b \frac{1}{m_a^2-s_1} \frac{1}{m_b^2-s_2}
\end{equation}
\begin{equation}
        \feynpdf[120pt]{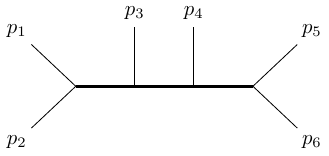}=\sum_{abc} \frac{g_a g_{ab} g_{bc} g_c }{2}\frac{1}{m_a^2-s_1} \frac{1}{m_b^2-s_2} \frac{1}{m_c^2-s_3}
\end{equation}
\begin{equation}
        \feynpdf[110pt]{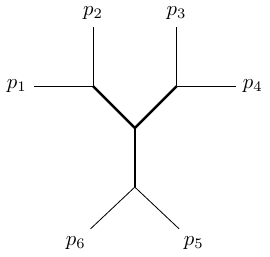}=\sum_{abc} \frac{g_a g_b g_c g_{abc}}{3!} \frac{1}{m_a^2 - s_1} \frac{1}{m_b^2 - t_{34}} \frac{1}{m_c^2-s_3}
\end{equation}
    We again expand the result in the heavy-mass limit and collect the resulting contributions. The independent polynomials in which we expand $\mathcal{A}_{6\phi}$ are all listed in appendix \ref{appendix explicit polynomials}, while their associated coefficients can be found in appendix \ref{appendix coefficients uv}. The additional factors of $1/2$ are, again, chosen to avoid double counting terms equivalent modulo momentum conservation.
    \begin{eqnarray}\label{eq 6point expanded}
        \mathcal{A}_{6\phi}=& \gamma^{(0)} + \gamma^{(2)} p^{(2)} + \gamma_A^{(3)} p_A^{(3)} + \gamma_B^{(3)} p_B^{(3)} + \gamma_A^{(4)} p_A^{(4)} + \gamma_B^{(4)} p_B^{(4)} +\gamma_C^{(4)} p_C^{(4)}+\gamma_D^{(4)} p_D^{(4)} \nonumber \\
        &+ \gamma_A^{(5)} p_A^{(5)} +\gamma_B^{(5)} p_B^{(5)} +\gamma_C^{(5)} p_C^{(5)} + \gamma_D^{(5)} p_D^{(5)}+ \gamma_E^{(5)} p_E^{(5)}+\gamma_F^{(5)} p_F^{(5)} \nonumber \\
        &+\gamma_A^{(6)} p_A^{(6)} + \gamma_B^{(6)} p_B^{(6)} +\gamma_C^{(6)} p_C^{(6)} + \gamma_D^{(6)} p_D^{(6)}+ \gamma_E^{(6)} p_E^{(6)} + \gamma_F^{(6)} p_F^{(6)} + \gamma_G^{(6)} p_G^{(6)} \nonumber \\
        &+\gamma_H^{(6)} p_H^{(6)} + \gamma_I^{(6)} p_I^{(6)} +\gamma_J^{(6)} p_J^{(6)}+ {\cal O}(s_i,t_{ij})^7 .
    \end{eqnarray}
    The sector relevant for our purposes is given in \eqref{fig coefficients and polynomials 6point}. These are all contributions independent of the cubic coupling $g_{abc}$, thereby circumventing the fate outlined in section \ref{sect obstruction gabc}. Our goal lies in finding exact relations among these coefficients.
    \begin{eqnarray}\label{fig coefficients and polynomials 6point}
        \gamma_D^{(4)}&=&\sum_a \frac{\widetilde{\lambda}_a^2}{m_a^{10}}\,, \qquad      \gamma_E^{(5)}= \sum_a \frac{\widetilde{\lambda}_a^2}{m_a^{12}}\, , \qquad
        \gamma_F^{(5)}=-\sum_{ab} \frac{g_a g_{ab} \widetilde{\lambda}_b}{m_a^6 m_b^8}\, , \qquad
        \gamma_G^{(6)}= -\sum_{ab} \frac{g_a g_{ab}\widetilde{\lambda}_b}{m_a^8 m_b^8} \nn \\
        \gamma_H^{(6)}&=&-\sum_{ab} \frac{g_a g_{ab}\widetilde{\lambda}_b}{m_a^6 m_b^{10}} \, , \qquad
        \gamma_I^{(6)}=\sum_{abc} \frac{g_a g_{ab} g_{bc} g_c}{m_a^6 m_b^6 m_c^6}  \, , \qquad 
        \gamma_J^{(6)}=\sum_a \frac{\widetilde{\lambda}_a^2}{m_a^{14}}\nn \\
        p_D^{(4)}&=&\frac{1}{2}\sum_{i<j<k} \(-(p_i+p_j+p_k)^2\)^4 \, , \qquad 
        p_E^{(5)}=\frac{1}{2}\sum_{i<j<k} \(-(p_i+p_j+p_k)^2\)^5 \nn\\
        p_F^{(5)}&=&\sum_{\substack{i<j\\ k\neq i,j}} \(-(p_i +p_j)^2\)^2 \(-(p_i+p_j+p_k)^2\)^3 \nn \\
        p_G^{(6)}&=&\sum_{\substack{i<j\\ k\neq i,j}} \(-(p_i +p_j)^2\)^3 \(-(p_i+p_j+p_k)^2\)^3\nn\\
        p_H^{(6)}&=&\sum_{\substack{i<j\\ k\neq i,j}} \(-(p_i +p_j)^2\)^2 \(-(p_i+p_j+p_k)^2\)^4\, , \qquad
        p_J^{(6)}=\frac{1}{2} \sum_{i<j<k} \(-(p_i+p_j+p_k)^2\)^6\nn\\
        p_I^{(6)}&=&\frac{1}{2}\sum_{\substack{i<j\\ k\neq i,j\\ l\neq i,j,k}} \(-(p_i +p_j)^2\)^2 \(-(p_i+p_j+p_k)^2\)^2 \(-(p_i+p_j+p_k+p_l)^2\)^2
    \end{eqnarray}
    
\subsection{First Attempt to Six-Point Positivity}\label{sect bounds 6point}
    Let us now look for positivity statements for the six-point amplitude by combining the expansion coefficients of $\mathcal{A}_{6\phi}$. We stress that at this point we have not yet enforced \eqref{eq gram constraint} on the kinematic polynomials, leaving an implicit relation between the kinematic variables \eqref{eq gen mandel def} in which the amplitude has been expanded.  
    
    As noted earlier, even to have the chance of finding a definite statement between the Wilson coefficients in \eqref{eq 6point expanded}, we must consider terms without any contribution of the sign-indefinite couplings $g_{abc}$. From the list given in figure \ref{fig coefficients and polynomials 6point}, we see that the first time such a term arises is at fourth order in the derivative expansion, which is manifestly positive 
    \begin{equation}\label{eq 6point positive 4th}
        \gamma_D^{(4)}=\sum_a \frac{\widetilde{\lambda}_a^2}{m_a^{10}}= \sum_a \frac{1}{m_a^{10}} \( \lambda_a -\sum_b \frac{g_{ab}g_b}{m_b^4} (m_a^2+3m_b^2)\)^2\geq0.
    \end{equation}
    Analogous coefficients arise at every order we evaluated. At the fifth and sixth order in the expansion, these are 
    \begin{equation}\label{eq 6point higher order pos}
    \begin{aligned}
        \gamma_E^{(5)}=&\sum_a \frac{\widetilde{\lambda}_a^2}{m_a^{12}},\hspace{1.5cm}
        \gamma_J^{(6)}=\sum_a \frac{\widetilde{\lambda}_a^2}{m_a^{14}} .
    \end{aligned}
    \end{equation}
    It is remarkable how similarly this particular set behaves to the four-point Wilson coefficients. Their associated kinematic polynomial corresponds to adding three external momenta raised to the corresponding power, as is apparent from figure \ref{fig coefficients and polynomials 6point}. Within this class, any such higher-order interaction is bounded in terms of $\gamma^{(4)}_D$ 
    \begin{equation}
    \begin{aligned}\label{hierarchy gamma D}
       0\leq \gamma_E^{(5)}=& \sum_a \frac{\widetilde{\lambda}_a^2}{m_a^{12}}\leq \frac{1}{\Lambda^{2}} \sum_a \frac{\widetilde{\lambda}_a^2}{m_a^{10}} =\frac{\gamma_D^{(4)}}{\Lambda^{2}},\\
       0\leq \gamma_J^{(6)}=& \sum_a \frac{\widetilde{\lambda}_a^2}{m_a^{14}}\leq \frac{1}{\Lambda^{4}} \sum_a \frac{\widetilde{\lambda}_a^2}{m_a^{10}} =\frac{\gamma_D^{(4)}}{\Lambda^{4}}.
    \end{aligned}
    \end{equation}
    We expect that analogous Wilson coefficients also arise at all higher orders when writing $\mathcal{A}_{6\phi}$ in an appropriate basis of crossing symmetric polynomials.
    Another manifestly positive Wilson coefficient is
    \begin{equation}\label{eq 6point positive 6th}
        \gamma_I^{(6)}= \sum_{a} \frac{1}{m_a^6} \( \sum_b\frac{g_{ab} g_b}{m_b^6}\)^2\geq 0.
    \end{equation}
    While we only expanded $\mathcal{A}_{6\phi}$ up to sixth order in $\frac{\Box}{m_a^2}$, it is possible that a similar hierarchy to \eqref{hierarchy gamma D} in terms of $\gamma_I^{(6)}$ arises at higher orders in the expansion. The two manifestly positive contributions $\gamma_I^{(6)}$ and $\gamma_J^{(6)}$ can also be combined with $\gamma_H^{(6)}$, where for an arbitrary real number $c$
    \begin{equation}\label{eq 6point positive mix}
    \begin{aligned}
        2 c \,\gamma_H^{(6)} + c^2 \gamma_I^{(6)} + \gamma_J^{(6)}=&-2 c\sum_{ab}\frac{g_a g_{ab}\widetilde{\lambda}_b}{m_a^6 m_b^{10}}+c^2 \sum_a \( \sum_b \frac{g_{ab} g_b }{m_a^3 m_b^6}\)^2 +\sum_a \frac{\widetilde{\lambda}_a^2}{m_a^{14}} \\
        =&\sum_a \(\frac{\widetilde{\lambda}_a}{m_a^7} - c \sum_b \frac{g_{ab}g_b}{m_a^3 m_b^6}\)^2\geq 0,
    \end{aligned}
    \end{equation}
    with the two instances $c\rightarrow 0, \infty$  capturing the positivity of $\gamma_J^{(6)}$ and $\gamma_I^{(6)}$ respectively. The manifest positivity for all values of $c$ will be crucial for stating bounds that survive imposing the Gram constraint \eqref{eq gram constraint}.
    Extremising over the free parameter $c$, one finds
    \begin{equation}
        c_{\text{extr}}=-\frac{\gamma_H^{(6)}}{\gamma_I^{(6)}}\,.
    \end{equation}
    This allows us to reduce the family \eqref{eq 6point positive mix} into the nonlinear bound 
    \begin{equation}\label{eq 6point cauchy 6th order}
        \gamma_I^{(6)} \gamma_J^{(6)}-\(\gamma_H^{(6)}\)^2= \sum_a \( \sum_b \frac{g_{ab} g_b }{m_a^3 m_b^6}\)^2 \cdot \sum_a \frac{\widetilde{\lambda}_a^2}{m_a^{14}}- \(\sum_{ab}\frac{g_a g_{ab}\widetilde{\lambda}_b}{m_a^6 m_b^{10}}\)^2\geq 0,
    \end{equation}
    which is an exact relation among expansion coefficients in the evaluated six-point amplitude at tree-level. It is captured by demanding positive semi-definiteness of the matrix
    \begin{equation}
        \begin{pmatrix}
            \gamma_J^{(6)} & \gamma_H^{(6)}\\
            \gamma_H^{(6)} & \gamma_I^{(6)}
        \end{pmatrix} \succeq 0.
    \end{equation}
    Statements that mix different orders in the derivative expansion can also be constructed, showcasing a rich structure arising in the sector unaffected by $g_{abc}$. Connecting different orders in the derivative expansion, we have
    \begin{eqnarray}\label{eq 6point cauchy general}
        &&\gamma_D^{(4)}\(2c\,\gamma_H^{(6)} +c^2 \gamma_I^{(6)} + \gamma_J^{(6)}\)-\( \gamma_E^{(5)} +c \, \gamma_F^{(5)}\)^2 \nn \\
        && =\sum_a \frac{\widetilde{\lambda}_a^2}{m_a^{10}} \cdot \sum_a \(\frac{\widetilde{\lambda}_a}{m_a^7} - c\sum_b \frac{g_{ab}g_b}{m_a^3 m_b^6}\)^2 -\(\sum_a \frac{\widetilde{\lambda}_a^2}{m_a^{12}}-c \sum_{ab} \frac{g_a g_{ab} \widetilde{\lambda}_b}{m_a^6 m_b^8}\)^2\geq 0,
    \end{eqnarray}
    for an arbitrary real parameter $c$. While one could also reduce this family of constraints to a single statement, it will turn out that the freedom in choosing $c$ will be useful once we impose the Gram constraint \eqref{eq gram constraint}. Even though the exact form of these expression is contingent on the specific model Lagrangian chosen in \eqref{lagrangian general renorm scalar}, it suggests not only that establishing new constraints from higher-point amplitudes is possible, but also where one might look for more general statements. For example the coefficients $\gamma_D^{(4)}$, $\gamma_E^{(5)}$, $\gamma_J^{(6)}$ that multiply symmetrised triplets of momenta are all positive and follow a hierarchy \eqref{hierarchy gamma D} as is familiar from four-point scattering. They can also be combined into exact Cauchy-Schwarz inequalities such as \eqref{eq 6point cauchy general} with $c=0$. Although this hierarchy is affected by the Gram constraint in $D=4$, as we discuss in the following section, it is still remarkable how similar the statements \eqref{eq 6point cauchy general} look to the well-established analogues found for four-particle scattering \eqref{eq bound 4point nonlin}.
    
    In addition to this, while the straightforward obstruction on bounding six-point amplitudes in the chosen model Lagrangian is already quite constraining, we note that the nonlinear bounds \eqref{eq 6point cauchy general} actually involve all but one Wilson coefficient that do not suffer the fate outlined in section \ref{sect obstruction gabc}. We leave it for further work to characterise this kinematic space in more detail.
    
\subsection{Enforcing the Gram Constraint}\label{sect imposing gram}
    While the conclusions drawn in the last subsection seem very promising, they are somewhat premature: as discussed in section \ref{sect kinematic variables}, the generalised Mandelstam variables \eqref{eq gen mandel def} are not all independent in $D=4$. Therefore, it must be checked whether the polynomials extracted in \eqref{eq 6point expanded} remain independent after enforcing the Gram constraint. It turns out that this is the case for all contributions at fourth and sixth order. However, imposing \eqref{eq gram constraint} implies a nontrivial relation between the polynomials $p_i^{(5)}$ at fifth order
    \begin{equation}\label{gram at fifth order}
    2 p_A^{(5)}+ 5 p_B^{(5)} -15 p_C^{(5)} +3 p_E^{(5)} - 5 p_F^{(5)} = 0.
    \end{equation}
    As the equation above involves both contributions $p_E^{(5)}$ and $p_F^{(5)}$ used in the nonlinear bounds in the last section, their associated Wilson coefficients are not accessible from expanding $\mathcal{A}_{6\phi}$ at low energies. If we want any term at fifth order in the derivative expansion that does not receive contributions from the coupling $g_{abc}$, we need to solve the above relation for either $p_E^{(5)}$ or $p_F^{(5)}$. Solving for $p_F^{(5)}$, the coefficient multiplying  $p_E^{(5)}$ becomes 
    \begin{equation}\label{eq shifted gamma}
        \widetilde{\gamma}_E^{(5)}=\gamma_E^{(5)} + \frac{3}{5} \gamma_F^{(5)} \, .
    \end{equation}
    The Wilson coefficients therefore lose the natural structure we observed in the last subsection at fifth order, since only the specific combination $\widetilde{\gamma}_E^{(5)}$ is accessible from a expanding an amplitude in $D=4$.
    Not only is the coefficient proportional to $p_E^{(5)}$ no longer manifestly positive, also the nonlinear bounds \eqref{eq 6point cauchy general} lose their validity in general. However, choosing $c=\frac{3}{5}$ accounts for the projection observed in \eqref{eq shifted gamma}, yielding a nonlinear six-particle bound that survives in $D=4$
    \begin{eqnarray}\label{eq cauchy mixed post gram}
        &&\gamma_D^{(4)} \(\gamma_J^{(6)}+\frac{6}{5}\gamma_H^{(6)}+\frac{9}{25}\gamma_I^{(6)}\)-\(\widetilde{\gamma}_E^{(5)} \)^2 \nn\\
        &&= \sum_a \frac{\widetilde{\lambda}_a^2}{m_a^{10}} \cdot \sum_a \(\frac{\widetilde{\lambda}_a}{m_a^7}- \frac{3}{5}\sum_b \frac{g_{ab}g_b}{m_a^3 m_b^6}\)^2 - \(\sum_a \frac{\widetilde{\lambda}_a^2}{m_a^{12}}- \frac{3}{5}\sum_{ab} \frac{g_a g_{ab} \widetilde{\lambda}_b}{m_a^6 m_b^8}\)^2\geq 0.
    \end{eqnarray}
    One may ask for an intuitive reason why imposing the Gram constraint \eqref{eq gram constraint} only affects terms at fifth order in the derivative expansion. As \eqref{eq gram constraint} is a quintic constraint in Mandelstam variables, the independence of lower-order contributions is not impacted by its presence. The fact that it does affect our conclusions drawn at fifth order shows that the specific combination of the $p_i^{(5)}$ stated is proportional to the Gram determinant. The possible analogue for sextic polynomials is that they combine into something proportional to the left-hand side of \eqref{eq gram constraint}. One must therefore check whether some combination of the $p_i^{(6)}$ results into the Gram determinant times an arbitrary linear polynomial. But as any crossing symmetric linear combination of the variables \eqref{eq gen mandel def} vanishes, imposing the Gram constraint does not affect the terms at sixth order. It is therefore expected that at even higher orders in the derivative expansion the Gram constraint would again relate otherwise independent polynomials. 

    Let us also briefly mention that imposing \eqref{eq gram constraint} is kinematically equivalent to working in $D=4$. Considering amplitudes in higher-dimensional spacetimes with $D\ge 5$ similar issues only arise at higher multiplicity, thereby restoring the hierarchy observed in section \ref{sect bounds 6point} among the expansion coefficients in the three-particle exchange channel. One should note that the considered model \eqref{lagrangian general renorm scalar} does not remain perturbatively renormalisable in higher spacetime dimension. For $D=5,6$, in order for the interpretation that the derived statements arise from a specific UV-complete ansatz Lagrangian, one must therefore set all quartic couplings in \eqref{lagrangian general renorm scalar} to zero. With this caveat in mind, the stated bounds remain valid for this case. We leave it for further work to study the relation between the analytic structure of six-particle amplitudes and the corresponding spacetime dimension in more detail.

\subsection{Bounds combining different Multiplicity}\label{sect bounds different amps}
    In addition to the rich structure found in the six-point amplitude, we can also try to relate expansion coefficients of amplitudes with different multiplicity. Establishing such a relation is of great interest, as it would pave the way to bounding six-particle scattering in terms of lower-point amplitudes. One particularly interesting example connects the dominant term in $\mathcal{A}_{4\phi}= \alpha^{(0)} +\alpha^{(2)} k^{(2)}+...$ with $\gamma_D^{(4)}$, i.e. the lowest-order contribution to $\mathcal{A}_{6\phi}$ that does not suffer the fate outlined in \ref{sect obstruction gabc}. They combine to an exact Cauchy-Schwarz inequality with the expansion coefficient $\beta^{(3)}$ from the five-point amplitude 
    \begin{equation}\label{eq multiplicity bound lowest order}
        \alpha^{(2)} \cdot \gamma_D^{(4)} -\( \beta^{(3)} \)^2 = \sum_a \frac{g_a^2}{m_a^6} \cdot \sum_a \frac{\widetilde{\lambda}_a^2}{m_a^{10}} -\(\sum_a \frac{g_a \widetilde{\lambda}_a}{m_a^8}\)^2 \geq 0,
    \end{equation}
    which can also be phrased as positivity of the determinant of
    \begin{equation} \label{alphabeta1}
        \begin{pmatrix}
            \alpha^{(2)} & \beta^{(3)}\\
            \beta^{(3)} & \gamma_D^{(4)}
        \end{pmatrix} \succeq 0.
    \end{equation}
It is worth stressing that despite the appearances, these are different from the planar EFT bounds of
Ref.~\cite{Cheung:2025nhw}. This is made particularly transparent by taking the limit
$g_{ab}=0$, with $g_a$ and $\lambda_a$ kept non-zero. In this limit, the
maximal half-ladder residues with two heavy propagators at five points
and three heavy propagators at six points vanish because their interior cubic
vertices require coupling $g_{ab}$. Nevertheless, the above equation \eqref{alphabeta1} continues to give a non-trivial bound.
    
    Since the Gram relation \eqref{eq gram constraint} spoils the positivity of the relevant six-point Wilson coefficient at fifth order, no similar statement can be stated for $\widetilde{\gamma}_E^{(5)}$. At sixth order in the expansion, however, we find a family of constraints labelled by a real parameter $c$
    \begin{eqnarray}\label{eq multiplicity bound family}
        && \alpha^{(2)}\(2 c\, \gamma_H^{(6)}+ c^2 \gamma_I^{(6)}+\gamma_J^{(6)} \)-\( \beta_A^{(4)}+c\,\beta_B^{(4)}\)^2 \nn \\
        && =\sum_a \frac{g_a^2}{m_a^6} \cdot \sum_a \(\frac{\widetilde{\lambda}_a}{m_a^7} - c\sum_b \frac{g_{ab}g_b}{m_a^3 m_b^6}\)^2-\(\sum_a \frac{g_a \widetilde{\lambda}_a}{m_a^{10}}-c\sum_{ab}\frac{g_a g_{ab} g_b}{m_a^6 m_b^6}\)^2\geq0.
    \end{eqnarray}
    We again highlight the situations in which $c$ tends to zero or infinity, in which case the above reduces to
    \begin{eqnarray}
       &&\alpha^{(2)} \cdot \gamma_J^{(6)} -\( \beta_A^{(4)} \)^2 = \sum_a \frac{g_a^2}{m_a^6} \cdot \sum_a \frac{\widetilde{\lambda}_a^2}{m_a^{14}} -\(\sum_a \frac{g_a \widetilde{\lambda}_a}{m_a^{10}}\)^2 \geq 0,\\\label{eq multiplicity bound 1}
        &&\alpha^{(2)} \cdot \gamma_I^{(6)} - \(\beta_B^{(4)} \)^2= \sum_a \frac{g_a^2}{m_a^6} \cdot \sum_a \frac{1}{m_a^6} \( \sum_b \frac{g_{ab}g_b}{m_b^6}\)^2-\( \sum_{ab} \frac{g_a g_{ab} g_b}{m_a^6 m_b^6}\)^2\geq0.\label{eq multiplicity bound 2}
    \end{eqnarray}
    Being exact statements about certain expansion coefficients these inequalities can be used to constrain scattering amplitudes involving six $\phi$-particles in terms of lower-point amplitudes. In order to find the strongest statement implied by \eqref{eq multiplicity bound family}, we extremise over the free parameter $c$. Evaluating the derivative of the left side with respect to $c$, we find 
    \begin{equation}
        c_{\text{extr}}=\frac{\beta_A^{(4)} \beta_B^{(4)}-\alpha^{(2)}\gamma_H^{(6)}}{\alpha^{(2)} \gamma_I^{(6)}-\(\beta_B^{(4)}\)^2}\,.
    \end{equation}
    Plugging this back into \eqref{eq multiplicity bound family}, and using \eqref{eq multiplicity bound 2}, we find after some algebra that 
    \begin{equation}\label{extremal mixed bound}
        \alpha^{(2)} \(\gamma_I^{(6)} \, \gamma_J^{(6)}-\(\gamma_H^{(6)}\)^2 \)-\(\beta_A^{(4)}\)^2 \gamma_I^{(6)}-\(\beta_B^{(4)}\)^2 \gamma_J^{(6)} +2 \beta_A^{(4)} \beta_B^{(4)} \gamma_H^{(6)}\geq 0.
    \end{equation}
    The implications of \eqref{eq multiplicity bound family}, together with \eqref{eq 6point cauchy 6th order}, are therefore summarised by demanding positivity of the determinant and all principal minors of the matrix 
    \begin{equation}\label{bound matrix}
        \begin{pmatrix}
            \alpha^{(2)} & \beta_A^{(4)} & \beta_B^{(4)} \\
            \beta_A^{(4)} & \gamma_J^{(6)} & \gamma_H^{(6)}\\
            \beta_B^{(4)} &\gamma_H^{(6)} & \gamma_I^{(6)}
        \end{pmatrix} \succeq  0.
    \end{equation}
    We stress that both statements \eqref{eq multiplicity bound lowest order} and \eqref{bound matrix} remain valid after imposing the Gram constraint \eqref{eq gram constraint}.
    
    In a completely analogous way we can construct a second positive semi-definite matrix. For this, we note that
\begin{eqnarray}
&&
\alpha^{(2)}\left(
\gamma_D^{(4)}
+2d\,\widetilde{\gamma}_E^{(5)}
+d^2  \left(\gamma_J^{(6)}+\frac{6}{5}\gamma_H^{(6)}+\frac{9}{25}\gamma_I^{(6)}\right)\right)
-\left(
\beta^{(3)}
+d
\left(
\beta_A^{(4)}
+\frac{3}{5}\beta_B^{(4)}
\right)
\right)^2
\\ \nn
&&=
\sum_a\frac{g_a^2}{m_a^6}
\cdot
\sum_c
\left[
\frac{\widetilde{\lambda}_c}{m_c^5}
\left(
1+\frac{d}{m_c^2}
\right)
-\frac{3d}{5}
\sum_b
\frac{g_{cb}g_b}{m_c^3m_b^6}
\right]^2
-\left[ \sum_a
\frac{g_a\widetilde{\lambda}_a}{m_a^8}
\left(
1+\frac{d}{m_a^2}
\right)
-\frac{3d}{5}
\sum_{a,b}
\frac{g_ag_{ab}g_b}{m_a^6m_b^6}
\right]^2
\geq0 
\end{eqnarray}
    holds, where we make use of \eqref{eq 6point positive mix} for $c=\frac{3}{5}$. Again, for the extremal value of $d$
    \begin{equation}
        d_{\text{extr}}=\frac{\beta^{(3)} \(\beta_A^{(4)}+\frac{3}{5} \beta_B^{(4)}\)-\alpha^{(2)} \widetilde{\gamma}_E^{(5)}}{\alpha^{(2)} \(\gamma_J^{(6)}+\frac{6}{5} \gamma_H^{(6)} +\frac{9}{25} \gamma_I^{(6)}\) -\(\beta_A^{(4)}+\frac{3}{5} \beta_B^{(4)}\)^2}
    \end{equation}
    the structure of a determinant emerges. Together with Eq.~\eqref{alphabeta1},
Eq.~\eqref{eq cauchy mixed post gram}, and the special case
$c=3/5$ of Eq.~\eqref{eq multiplicity bound family}, which reproduce
the three $2\times2$ principal minors, the determinant inequality
establishes
     \begin{equation}\label{bound matrix2}
        \begin{pmatrix}
            \alpha^{(2)} & \beta^{(3)} & \beta_A^{(4)}+\frac{3}{5}\beta_B^{(4)}  \\
            \beta^{(3)} & \gamma_D^{(4)} & \widetilde{\gamma}_E^{(5)}\\
            \beta_A^{(4)}+\frac{3}{5}\beta_B^{(4)}  &\widetilde{\gamma}_E^{(5)} & \gamma_J^{(6)}+ \frac{6}{5}\gamma_H^{(6)}+\frac{9}{25}\gamma_I^{(6)}
        \end{pmatrix} \succeq  0.
    \end{equation}
   The above Gram-matrix structure bounds look similar to those of Ref.~\cite{Cheung:2026lpv}, however, the construction outlined there assumes planarity in order to relate low-energy coefficients to residues in a single factorisation channel. Our amplitudes include all crossed channels.

    We will discuss how we can use these together with the bounds stated in the last sections to bound interactions between six $\phi$-particles in terms of the four- and five-point amplitudes later. Before going there, we need to consider contributions arising from light exchanges.
    
\subsection{Contribution of Light Exchanges}\label{sect light exchanges uv}
    Alongside the amplitude contributions considered so far, we have to account for interactions mediated by light internal propagators. The key difference to the discussion above will be that these are not expanded in the low-energy limit, implying that diagrams yield both pole and analytic contributions. The question arises how to split these in an unambiguous way. To see this, consider the diagram\footnote{Note that as the cubic interactions do not include any derivatives, such a procedure is not needed for $\mathcal{A}_{4\phi}$.}
    \begin{equation}
    \feynpdf[120pt]{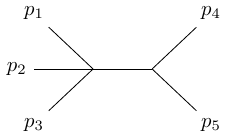}
        = \frac{\mathcal{F}(s_1,s_2,t_{23},...)}{s_2} \, ,
        \end{equation}
    contributing to scattering of five $\phi$-particles, where the internal propagator is now light. We want to write this as 
    \begin{equation}
        \frac{\mathcal{F}(s_1,s_2,t_{23},...)}{s_2}=\frac{\mathcal{F}(s_1,0,t_{23},...)}{s_2}+\frac{\mathcal{F}(s_1,s_2,t_{23},...)-\mathcal{F}(s_1,0,t_{23},...)}{s_2}
    \end{equation}
    where the second term will be analytic in $s_2$, as can easily be seen by expanding the numerator around $s_2=0$. For some exchange momentum represented by a combination of Mandelstams, say $s_1 - s_2 - t_{34}$, we need to choose whether we extract the analytic part by interpreting it as a pole in $s_1$, $s_2$ or $t_{34}$. While there are different choices possible, it is crucial that the prescription to pick out the analytic contribution is unambiguous and can be repeated in the EFT. For the five-point amplitude we use the prescription that:
    \begin{itemize}
        \item if a denominator only depends on a single kinematic variable, it has to go on-shell. Otherwise
        \item if the denominator involves $s_2$, interpret it as a $s_2$ pole. Otherwise,
        \item interpret it as a $t_{24}$ pole.
    \end{itemize}
    By explicitly writing down all possible internal momenta it can easily be checked that every ambiguous propagator involves either $s_2$ or $t_{24}$. This prescription yields an additional analytic contribution to the five-$\phi$  amplitude at each order in the derivative expansion. The fact that it is independent of the terms arising from heavy exchanges is obvious, as our prescription is not crossing symmetric. However, if there is more than one distinguishable light-exchange polynomial contributing, they could still combine into a crossing symmetric expression. 
    
    Next we repeat the same procedure for the six-point amplitude. The prescription we use for $\mathcal{A}_{6\phi}$ to resolve ambiguous poles is
    \begin{itemize}
         \item if a denominator only depends on a single kinematic variable, it goes on-shell. Otherwise
        \item if the denominator involves $s_2$, interpret it as a $s_2$ pole. Otherwise,
        \item if the denominator involves $t_{24}$, interpret it as a $t_{24}$ pole. Otherwise,
        \item interpret it as a $t_{35}$ pole.
    \end{itemize}
    There is an additional complication for six-point scattering here, as one has to consider diagrams with more than one light propagator. If there are such propagators in a single diagram, we apply this prescription sequentially in order to extract the fully analytic part. By this procedure, we can unambiguously separate the divergent piece and the analytic contribution. Consider, for example,
    \begin{equation}
        \feynpdf[120pt]{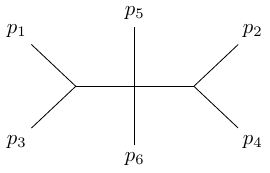}=\frac{\mathcal{F}(s_1,s_2,s_3,t_{23},t_{24},...)}{(-s_1+s_2-t_{23})(-t_{23}+t_{24}-t_{34})}.
    \end{equation}
    We extract the analytic part by first considering the pole $s_2\rightarrow s_1+t_{23}$. We then write 
    \begin{equation}
    \begin{aligned}
        \frac{\mathcal{F}(s_1,s_2,s_3,...)}{(-s_1+s_2-t_{23})(-t_{23}+t_{24}-t_{34})}=&\frac{\mathcal{F}(s_1,s_1+t_{23},s_3,...)}{(-s_1+s_2-t_{23})(-t_{23}+t_{24}-t_{34})}\\ 
        &+\underbrace{\frac{\mathcal{F}(s_1,s_2,s_3,...)-\mathcal{F}(s_1,s_1+t_{23},s_3,t_{23},t_{24},...)}{(-s_1+s_2-t_{23})(-t_{23}+t_{24}-t_{34})}}\\
        & \hspace{2cm} \equiv \frac{\widetilde{\mathcal{F}}(s_1,s_2,s_3,...)}{-t_{23}+t_{24}-t_{34}}        
    \end{aligned}
    \end{equation}
    with the second term being fully analytic in $s_2$. Next we account for the second pole by writing
    \begin{equation}
    \begin{aligned}
        &\frac{\widetilde{\mathcal{F}}(...,t_{23},t_{24},t_{25},...)}{-t_{23}+t_{24}-t_{34}}= \underbrace{\frac{\widetilde{\mathcal{F}}(...,t_{23},t_{24},t_{25},...)-\widetilde{\mathcal{F}}(...,t_{23},t_{23}+t_{34},t_{25},...)}{-t_{23}+t_{24}-t_{34}}}+\frac{\widetilde{\mathcal{F}}(...,t_{23},t_{23}+t_{34},t_{25},...)}{-t_{23}+t_{24}-t_{34}}.\\
        & \hspace{6cm} \equiv \widetilde{\widetilde{\mathcal{F}}}(...,t_{23},t_{24},t_{25},...)
    \end{aligned}
    \end{equation}
    The function $\widetilde{\widetilde{\mathcal{F}}}$ gives a fully analytic contribution to $\mathcal{A}_{6\phi}$. We note that the result following from this procedure is independent of the order of resolution, if each pole is associated with a single Mandelstam variable. For poles in the same variable, we decompose into partial fraction to treat them separately. For example:
    \begin{equation}
    \begin{aligned}
        \frac{\mathcal{F}(s_1,s_2,s_3,...)}{(-s_1+s_2-t_{23})(s_2-s_3-t_{45})}= \frac{1}{s_1+t_{23}-s_3-t_{45}} \(\frac{\mathcal{F}(s_1,s_2,s_3,...)}{s_2-s_1-t_{23}}-\frac{\mathcal{F}(s_1,s_2,s_3,...)}{s_2-s_3-t_{45}}\).
    \end{aligned}
    \end{equation}
    Its analytic contribution follows from extracting the corresponding terms from both $s_2$ poles separately, with their sum cancelling the spurious pole. 

    After isolating the analytic contributions obtained in the outlined manner, we project them onto the set of crossing symmetric polynomials that capture heavy exchanges. For more details on the implementations of this, we refer readers to appendix \ref{appendix light exchanges}. The most pressing question for our purposes is, whether terms obtained using the detailed prescription overlap with polynomials whose Wilson coefficients show up in the constructed positive matrices \eqref{bound matrix}, \eqref{bound matrix2}. Comparing the span of light-exchange polynomials to the span of heavy ones, we indeed observe some mixing. More precisely, the set formed by combining a basis of independent light-exchange polynomials with another basis of independent heavy-exchange polynomials will not itself be independent. However, while the two spaces are not disjoint, it turns out that at fourth, fifth and sixth order this intersection is each one-dimensional only. Using the basis of polynomials given in appendix \ref{appendix explicit polynomials}, the direction in which the mixing happens is given by $p_A^{(4)}$, $p_A^{(5)}$ and $p_A^{(6)}$. For more details on this we refer interested readers to appendix \ref{appendix light exchanges}. Crucially, at each order the mixing happens only in the sector receiving contributions from the sign-indefinite couplings $g_{abc}$ without affecting any of the terms related via our positivity statements \eqref{bound matrix}, \eqref{bound matrix2}.

    The bounds stated prior are therefore exact statements relating the full (connected, tree-level) four-, five- and six-point amplitude using the prescription just outlined. While this separation of light and heavy contributions is contingent on the chosen prescription, the underlying coupling-space inequalities hold independent of the prescription chosen. In another consistent subtraction scheme one must transform, or explicitly subtract, the corresponding light-exchange analytic contributions before applying the bounds stated in prior sections. 
    
\section{Bounding Six-Point Interactions in Single-Scalar EFT}\label{section eft}
    \subsection{Single-Scalar EFT}
    In order to impose the derived bounds on a single-scalar EFT to see their implication, we must first write such a theory down. For this we need to establish a basis of relevant EFT interactions. Due to the freedom of field redefinitions, there is an inherent ambiguity in choosing a minimal set of independent terms at any derivative order. 
    To capture the $k$-th order contribution to $n$-particle scattering, we need a minimal set of interactions such that every possible contraction of $k$ derivatives among $n$ field-operators can be expressed in terms of this list. This is comparable to choosing a basis for the crossing symmetric polynomials that span the amplitudes evaluated in section \ref{sect expand uv}. However, it is not clear a priori how long this list has to be, i.e. how many inequivalent interactions there are modulo field redefinitions and partial integration. For this reason, we started by writing down all possible contractions of the $k$ derivatives between $n$ fields. We then used partial integration and field redefinitions in order to reduce this set as much as possible, arriving at the lists given in appendix \ref{appendix eft interactions}, where we define the field-operator $\phi$ such that there are no cubic derivative interactions.\\
    In order to apply the bounds derived in section \ref{sect bounds 6point} and \ref{sect bounds different amps}, we need to evaluate contributions quadratic in Mandelstam variables in $\mathcal{A}_{4\phi}^{\text{EFT}}$, the cubic and quartic terms in $\mathcal{A}_{5\phi}^{\text{EFT}}$ as well as quartic, quintic and sextic contributions in $\mathcal{A}_{6\phi}^{\text{EFT}}$. The full set of derivative interactions to evaluate each amplitude up to ${\cal O}(\Lambda^{-14})$ in the EFT can be found in appendix \ref{appendix eft interactions}. In addition to evaluating these contact terms, one must also apply the prescription outlined in section \ref{sect light exchanges uv} to exchange diagrams. Due to our choice to absorb cubic derivative interactions into the field operator, this only has to be done for $\mathcal{A}_{5\phi}$ and $\mathcal{A}_{6\phi}$. For five-point scattering, the only relevant exchange diagrams are of the form
    \begin{equation*}
        \feynpdf[100pt]{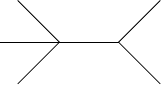}
    \end{equation*}
    with the quartic vertex including eight/ten derivatives for the result to become a cubic/quartic polynomial in Mandelstams. 
    In line with the UV calculation, we check for independence between the exchange sector and the contact interactions via the same method outlined in appendix \ref{appendix light exchanges}. \\
    For $\mathcal{A}_{6\phi}$ there are four relevant classes of exchange diagrams.
    \begin{equation*}
    \begin{aligned}
    &\feynpdf[100pt]{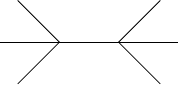} \hspace{2cm} \feynpdf[80pt]{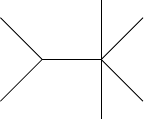}\\
    &\feynpdf[100pt]{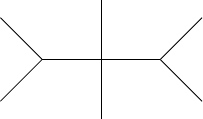}\hspace{2cm} \feynpdf[120pt]{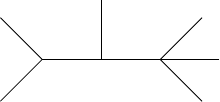}    
    \end{aligned}
    \end{equation*}
    Again, applying the prescription outlined in section \ref{sect light exchanges uv}, we extract their analytic contributions and compare the result terms with contact terms at each order in the derivative expansion, according to appendix \ref{appendix light exchanges}. This allows us to construct a maximal set of distinguishable polynomials $\mathcal{P}$ in which we can expand the analytic part of the amplitude.

    \subsection{EFT Matching}\label{sect eval eft}
    Using the set of polynomials $\mathcal{P}$ constructed in the last section as a basis, we can now compare the EFT with our UV calculation in the limit $\frac{\Box}{m_a^2}\ll1$. For this, the UV polynomials $k^{(n)}$, $q^{(n)}$ and $p^{(n)}$ in which $\mathcal{A}_{4\phi}$, $\mathcal{A}_{5\phi}$ and $\mathcal{A}_{6\phi}$ were expanded in section \ref{sect expand uv}, as well as the corresponding light-exchange contributions, get projected onto $\mathcal{P}$. The resulting coefficients then make up a vector that specifies each UV polynomial in the space of EFT contributions involving $k$ derivatives and $n$ fields. In order to find EFT analogues of the bounds \eqref{bound matrix}, \eqref{bound matrix2} we then have to invert this map. It turns out that this is straightforward for the UV and EFT basis considered here, since the original map projecting the UV into EFT space is already nearly diagonal. There is some mixing observed between the light- and heavy-exchange contributions, as well as non-diagonal matching relations. However, the relevant terms for our endeavours, which are the polynomials whose Wilson coefficients show up in \eqref{bound matrix} and \eqref{bound matrix2}, each manifest as exactly one contact term in the EFT.

    As all interactions among four field-operators with four derivatives contracted among them are equivalent, we choose $\frac{a^{(2)}}{8} (\partial \phi)^2(\partial \phi)^2$. The associated amplitude contribution to $\mathcal{A}_{4\phi}^{\text{EFT}}$ can be read off after changing to momentum space. Since we work in a basis with no cubic derivative interactions the matching here is trivial since both the span of UV and EFT polynomials is one-dimensional. We find $a^{(2)} \leftrightarrow 4 \,\alpha^{(2)}$ where the factor of 4 follows from converting the contracted derivatives into boxes, as $(p+q)^2 = 2 \, p\cdot q$ for massless on-shell momenta $p$, $q$.
    For interactions involving five $\phi$-operators, we find that all interactions involving six derivatives are equivalent. The one we include in our basis is $\frac{b^{(3)}}{4}\phi (\partial\phi)^2 (\partial^{(2)} \phi)^2$. Matching to our UV model yields $b^{(3)} \leftrightarrow 8\, \beta^{(3)}$. Similarly, the single third order light-exchange contribution $q_l^{(3)}$ which is stated in appendix \ref{appendix light exchanges} is proportional to the analogously constructed single light-exchange term in the EFT at this order. At quartic order in Mandelstam variables two inequivalent terms arise, which are chosen to be $\frac{b_A^{(4)}}{4}\phi (\partial\phi)^2 (\partial^{(3)} \phi)^2$ and $\frac{b_B^{(4)}}{8}\phi (\partial^{(2)} \phi)^2 (\partial^{(2)} \phi)^2$. The two interaction terms are in one-to-one correspondence with the crossing symmetric polynomials stated in section \ref{sect expand uv}, such that $b_A^{(4)} \leftrightarrow 16\, \beta_A^{(4)}$, $b_B^{(4)} \leftrightarrow 16\, \beta_B^{(4)}$. The light-exchange contributions obtained in UV and EFT are also proportional to each other. 
    
    Repeating the same for six particles, we have four distinct interactions involving eight derivatives, five terms involving ten and thirteen contributions with twelve derivatives in total.\footnote{As was the case in our UV model, imposing the Gram constraint induces one constraint among quintic terms, while leaving quartic and sextic contributions unaffected. This reduces the number of independent terms in $D=4$ involving ten derivatives from six to five.} The basis of interactions at every order is given in appendix \ref{appendix eft interactions}. 
    Comparing these contributions with the expanded six-point amplitude \eqref{eq 6point expanded}, we find the EFT analogues of the couplings constrained by the statements derived in sections \ref{sect bounds 6point}, \ref{sect imposing gram} and \ref{sect bounds different amps}. The amplitude matching gives $c_D^{(4)}\leftrightarrow\gamma_D^{(4)}$, $c_E^{(5)}\leftrightarrow \widetilde{\gamma}_E^{(5)}$, $c_H^{(6)} \leftrightarrow 4 \, \gamma_H^{(6)}$, $c_I^{(6)} \leftrightarrow 16 \, \gamma_I^{(6)}$ and $c_J^{(6)}\leftrightarrow \gamma_J^{(6)}$.
    One can equally obtain the analogues of all other UV contributions in the EFT, with the basis of derivative interactions in appendix \ref{appendix eft interactions} chosen to make the mapping diagonal. However, since the coefficients stated above are the only ones necessary to translate the obtained bounds \eqref{bound matrix}, \eqref{bound matrix2} into the EFT, we refrain from writing other matching conditions here. Note that the prefactors necessary for the translation stem from rewriting all derivatives in these interactions into boxes. It should be mentioned that the chosen ansatz Lagrangian \eqref{Lagrangian UV only phi} does not generate the full set of independent derivative interactions involving twelve derivatives between six fields. It is not surprising that a specific ansatz such as \eqref{lagrangian general renorm scalar} does not populate the whole landscape of derivative interactions. This does not affect the bounds below, which follow from the image of the specific UV ansatz in EFT coefficient space. Specifically, the coefficients $\gamma_H^{(6)}$, $\gamma_I^{(6)}$, $\gamma_J^{(6)}$ involved in the derived bounds map to $c_H^{(6)}$, $c_I^{(6)}$, $c_J^{(6)}$. Having additional independent derivative interactions does not affect this map, leaving the drawn conclusions intact.
    
\subsection{Applying Bounds to EFT}
    Let us start by discussing the physical bounds that apply purely to the six-point amplitude. The lowest-order statement on $\mathcal{A}_{6\phi}^{\text{EFT}}$ is at ${\cal O}(\Lambda^{-{10}})$ given in \eqref{eq 6point positive 4th}. Matching with the EFT, the analogue statement is 
    \begin{equation}
        c_D^{(4)}\geq 0.
    \end{equation}
    Applying the observed hierarchy in \eqref{hierarchy gamma D} we obtain a compact bound for the strength of $\frac{c_J^{(6)}}{2 \, 3!\, 3!} \phi^3 \Box^6 \phi^3$ in
    \begin{equation}
        \frac{c_D^{(4)}}{\Lambda^4}\geq c_J^{(6)}\geq 0.
    \end{equation}
    As discussed in section \ref{sect imposing gram}, an analogue expression at quintic order is spoiled by the Gram constraint in $D=4$.
    A nonlinear bound obtained that involves only the six-point amplitude is \eqref{eq cauchy mixed post gram}. Following the EFT matching procedure discussed in section \ref{sect eval eft}, it implies the constraint
    \begin{equation}
        c_D^{(4)} \(\frac{3}{10} c_H^{(6)} + \frac{9}{400}c_I^{(6)}+c_J^{(6)} \) -\(c_E^{(5)}\)^2\geq 0,
    \end{equation}
    on the six-particle Wilson coefficients. This bound shows up as a particular minor of the positive semi-definite matrix \eqref{bound matrix2}  that manifests in the EFT as 
\begin{equation}\label{bound matrix2 EFT}
\begin{pmatrix}
4a^{(2)}
&
b^{(3)}
&
b_A^{(4)}+\dfrac{3}{5}b_B^{(4)}
\\
b^{(3)}
&
4c_D^{(4)}
&
8c_E^{(5)}
\\
b_A^{(4)}+\dfrac{3}{5}b_B^{(4)}
&
8c_E^{(5)}
&
16c_J^{(6)}
+\dfrac{24}{5}c_H^{(6)}
+\dfrac{9}{25}c_I^{(6)}
\end{pmatrix}
\succeq 0,
\end{equation}
    with the other positive matrix \eqref{bound matrix} given by
     \begin{equation}
       \begin{pmatrix}
            4 a^{(2)} & b_A^{(4)} &  b_B^{(4)} \\
            b_A^{(4)} & 16 c_J^{(6)} & 4 c_H^{(6)}\\
            b_B^{(4)} & 4 c_H^{(6)} & c_I^{(6)}
        \end{pmatrix} \succeq  0.\label{EFT multiplicity family}
    \end{equation}
    Since all principal minors of these matrices are nonnegative, they capture several bounds stated on their own at earlier stages in the paper, such as the non-linear pure six-particle bound \eqref{eq 6point cauchy 6th order}, for example. The stated prefactors that arise here are the result of the basis of derivative interactions chosen and could be easily removed by a different convention. In particular, converting all derivatives in the Appendix \ref{appendix eft interactions} into boxes would recover a form without any additional prefactors multiplying the different Wilson coefficients compared to \eqref{bound matrix} and \eqref{bound matrix2}.
   
    The stated inequalities are novel constraints on EFT interactions between six particles, using information about scattering at lower expansion order and multiplicity. Although a violation of these statements only implies an inconsistency with a UV completion \eqref{lagrangian general renorm scalar}, they show at least in this class that a very rich structure emerges in the kinematic sector independent of the sign-indefinite couplings outlined in section \ref{sect obstruction gabc}.

\section{Conclusions}\label{section conclusion}
    In this paper, we have derived a set of non-planar positivity bounds constraining EFT interactions between six scalar particles. To achieve this, we studied amplitudes in a specific model class, namely a general renormalisable scalar field theory in $D=4$. We constructed a set of sign-definite combinations of low-energy expansion coefficients of the four-, five- and six-point amplitudes at tree-level. While a violation of these bounds only implies the inability of an EFT to be UV completed into a Lagrangian as in \eqref{lagrangian general renorm scalar}, they motivate what structures may emerge more generally. Equally important, they help us to get clearer expectations of what seems unrealistic in a more general setting. As outlined in section \ref{sect obstruction gabc}, one of the key insights from studying our model is that there are natural interactions for which establishing any kind of positivity statement seems implausible. These difficulties already suggest that higher-point positivity is probably less universal when compared to four-point scattering.
    
    We saw that written in terms of the nine generalised Mandelstam variables \eqref{eq gen mandel def}, there is a sector in $\mathcal{A}_{6\phi}$ that does not receive contributions from these sign-indefinite couplings mentioned in section \ref{sect obstruction gabc}. All constraints obtained relate these terms, which interestingly show a positivity structure similar to four-point scattering. 
    In the context of our model Lagrangian \eqref{lagrangian general renorm scalar}, the four-point amplitude can be written using the kinematic moments 
    \begin{equation}
        \frac{1}{2} \sum_{i<j} \(-(p_i+p_j)^2\)^n \, ,
    \end{equation}
    with positive Wilson coefficients proportional to $g_a^2$ for $n\geq 2$. In the six-point heavy-exchange sector we find an analogous sequence 
    \begin{equation}
        \frac{1}{2}\sum_{i<j<k}\(-(p_i+p_j+p_k)^2\)^n \, ,
    \end{equation}
    with their positive coefficients proportional to $\widetilde{\lambda}_a^2$ defined in \eqref{eq def shifted lambda} for all considered orders $n \geq 6$. This particular three-particle exchange sector therefore allows for the construction of moment inequalities as familiar from four-particle scattering. These structures seem interesting candidates to study in a more general setting, although imposing the four-dimensional Gram condition \eqref{eq gram constraint} spoils the independence of the necessary coefficients at fifth order and most likely also at and beyond seventh order. It may be instructive to study the analytic structure of the amplitude before imposing the four-dimensional Gram relation and thinking of the physical amplitude as a Gram hypersurface in this space. Working with nine unconstrained six-point Mandelstam variables is kinematically equivalent to assuming five independent momenta, i.e. working in $D\geq 5$. More work is needed to check whether this approach aids our understanding of physical amplitudes sufficiently to justify such an extension.

    Returning to $D=4$, while a lot of the natural structure obtained seems to be lost after imposing the constraint \eqref{eq gram constraint},  we were still able to construct Cauchy-Schwarz relations among the physical expansion coefficients in the six-point amplitude in \eqref{eq 6point cauchy 6th order} and \eqref{eq cauchy mixed post gram}. Their structural similarity to the four-point bounds \eqref{eq bound 4point nonlin}, which are firmly established as implications of unitarity, causality and the amplitude's Regge behaviour \cite{Tolley:2020gtv,Caron-Huot:2020cmc}, suggests that they may also be remnants of more general bounds that follow from analogue properties of six-point amplitudes.

    In addition to deriving bounds for pure six-particle scattering, we also find relations between terms of different multiplicity. It is particularly interesting that the statement \eqref{eq intro result lowest order} obtained bounds the lowest-order expression that does not receive contributions from the sign-indefinite interactions described in section \ref{sect obstruction gabc}. It allows us to constrain a specific six-particle interaction in terms of information about lower-point scattering data. In this sense, they seem similar to the bounds derived in \cite{Cheung:2025nhw}, which discusses relations based on connecting residues in a specific kinematic channel at each multiplicity. Our work suggests that such statements may also hold outside the planar limit, where one cannot directly pick out residues of individual channels from the EFT-amplitude. At higher orders in the derivative expansion, we constructed positive matrices \eqref{result intro mixed orders}, \eqref{result intro higher order} that connect EFT interactions of different multiplicity. This seems to be aligned with other structures that often arise when bounding EFT interactions, suggesting that they could be manifestations of some more general statements rather than being a fluke of the considered model. The fact that all obtained statements that connect different multiplicities can be phrased in terms of positive semi-definite matrices suggests that they are manifestations of an unknown underlying identity, possibly a dispersion relation. 

\acknowledgments
The authors thank Peter Petrov for useful discussion in the early stages of this project. AJT is supported by the STFC Consolidated Grant ST/X000575/1. SK is supported by an STFC studentship. The algebraic computations in this work were carried out via Wolfram Mathematica. The authors also acknowledge the use of GPT-5.6 Sol as a supportive tool for debugging the associated computational scripts.

\newpage

\appendix
\section{Polynomials in Five- and Six-point amplitude}\label{appendix explicit polynomials}
    We list the choices of crossing symmetric polynomials in which we expand the five- and six-point amplitudes. Although our calculations were performed using generalised Mandelstam variables \eqref{eq gen mandel def}, we give them here in terms of momenta. This makes the expressions much more manageable and should help the reader to distinguish between different contributions. For five particle scattering, we also state the result in terms of the generalised Mandelstam variables \eqref{eq gen mandel def}. 

\renewcommand{\arraystretch}{1.5}
    \subsection{Five-Particle Scattering}
        $$\begin{array}{ll}
            q^{(2)}=\sum_{i<j} (-(p_i +p_j)^2)^2=&
        s_1^2 + (-s_1 + s_2 - t_{23})^2 + (- s_2 + t_{23} - t_{24})^2 + t_{24}^2 + t_{23}^2 \\
        &+ (- t_{23} + t_{24} - t_{34})^2 + (- s_1 - t_{24} + t_{34})^2 + t_{34}^2 + (s_1 - s_2 - t_{34})^2 +  s_2^2\\
     q^{(3)}= \sum_{i<j} (-(p_i +p_j)^2)^3=&
        s_1^3 + (-s_1 + s_2 - t_{23})^3 + (- s_2 + t_{23} - t_{24})^3 + t_{24}^3 + t_{23}^3 \\
        &+ (- t_{23} + t_{24} - t_{34})^3 + (- s_1 - t_{24} + t_{34})^3 + t_{34}^3 + (s_1 - s_2 - t_{34})^3 +  s_2^3\\
     q_A^{(4)}=\sum_{i<j} \(-(p_i +p_j)^2\)^4 =&
        s_1^4 + (-s_1 + s_2 - t_{23})^4 + (- s_2 + t_{23} - t_{24})^4 + t_{24}^4 + t_{23}^4 \\
        &+ (- t_{23} + t_{24} - t_{34})^4 + (- s_1 - t_{24} + t_{34})^4 + t_{34}^4 + (s_1 - s_2 - t_{34})^4 +  s_2^4\\
    q_B^{(4)}=\frac{1}{2}\sum_{i<j,k\neq i,j} \(-(p_i +p_j)^2\)^2&\(-(p_i+p_j+p_k)^2\)^2\\
       \hfill =& \frac{1}{2} \biggl( s_1^2 \(s_2^2 + t_{34}^2 + (-s_1 + s_2 + t_{34})^2\)\\
        &+(-s_1 + s_2 - t_{23})^2 \(s_2^2 + (s_1 + t_{24} - t_{34})^2 + (t_{23} - t_{24} + t_{34})^2\)\\
        &+ (- s_2 + t_{23} - t_{24})^2 \(t_{23}^2 + (s_1 + t_{24} - t_{34})^2 + (-s_1 + s_2 + t_{34})^2\)\\
        &+t_{24}^2 \(t_{23}^2 + t_{34}^2 + (t_{23} - t_{24} + t_{34})^2\)+ t_{23}^2 \(s_2^2 + t_{24}^2 + (s_2 - t_{23} + t_{24})^2\)\\
        &+ (- t_{23} + t_{24} - t_{34})^2 \((s_1 - s_2 + t_{23})^2 + t_{24}^2 + (-s_1 + s_2 + t_{34})^2\)\\
        &+(- s_1 - t_{24} + t_{34})^2 \((s_1 - s_2 + t_{23})^2 + (s_2 - t_{23} + t_{24})^2 + t_{34}^2\)\\
        &+t_{34}^2 \(s_1^2 + t_{24}^2 + (s_1 + t_{24} - t_{34})^2\)\\
        &+ (s_1 - s_2 - t_{34})^2 \(s_1^2 + (s_2 - t_{23} + t_{24})^2 + (t_{23} - t_{24} + t_{34})^2\)\\
        &+s_2^2 \(s_1^2 + t_{23}^2 + (s_1 - s_2 + t_{23})^2\)\biggr)
        \end{array}
        $$
        \vfill
    \subsection{Six-Particle Scattering}

    $$
   \begin{array}{ll}
        p^{(2)}=& \sum_{i<j} \(-(p_i +p_j)^2\)^2\\
        p_A^{(3)}=& \sum_{i<j} \(-(p_i +p_j)^2\)^3\\
        p_B^{(3)}=&\frac{1}{2} \sum_{i<j<k} \((-(p_i +p_j+p_k)^2\)^3\\  
        p_A^{(4)}=&\sum_{i<j} \(-(p_i +p_j)^2\)^4\\ 
        p_B^{(4)}=&\frac{1}{2} \sum_{i<j; k<l; i,j\neq k,l} \(-(p_i +p_j)^2\)^2\(-(p_k+p_l)^2\)^2\\
        p_C^{(4)}=&\sum_{i<j; k\neq i,j} \(-(p_i +p_j)^2\)^2\(-(p_i+p_j+p_k)^2\)^2\\
        p_D^{(4)}=&\frac{1}{2}\sum_{i<j<k} \(-(p_i+p_j+p_k)^2\)^4\\
        p_A^{(5)}=& \sum_{i<j} \(-(p_i +p_j)^2\)^5\\
        p_B^{(5)}=&\sum_{i<j;k<l;i,j\neq k,l} \(-(p_i +p_j)^2\)^3 \(-(p_k+p_l)^2\)^2\\
        p_C^{(5)}=&\frac{1}{2} \sum_{i<j;k<l;m<n;i,j\neq k,l\neq m,n} \(-(p_i +p_j)^2\)^2\(-(p_k+p_l)^2\)^2 \(-(p_m+p_n)^2\)\\
        p_D^{(5)}=& \sum_{i<j;k\neq i,j} \(-(p_i +p_j)^2\)^3 \(-(p_i+p_j+p_k)^2\)^2\\
        p_E^{(5)}=& \frac{1}{2}\sum_{i<j<k} \(-(p_i+p_j+p_k)^2\)^5\\
        p_F^{(5)}=& \sum_{i<j;k\neq i,j} \(-(p_i +p_j)^2\)^2 \(-(p_i+p_j+p_k)^2\)^3\\
        p_A^{(6)}=& \sum_{i<j} \(-(p_i +p_j)^2\)^6\\
        p_B^{(6)}=& \sum_{i<j;k<l;i,j\neq k,l} \(-(p_i +p_j)^2\)^4 \(-(p_k+p_l)^2\)^2\\
        p_C^{(6)}=& \frac{1}{2} \sum_{i<j;k<l;i,j\neq k,l} \(-(p_i +p_j)^2\)^3 \(-(p_k+p_l)^2\)^3\\
        p_D^{(6)}=& \sum_{i<j;k<l;m<n;i,j\neq k,l\neq m,n} \(-(p_i +p_j)^2\)^3 \(-(p_k+p_l)^2\)^2 \(-(p_m+p_n)^2\)\\
        p_E^{(6)}=&\frac{1}{6} \sum_{i<j;k<l;m<n;i,j\neq k,l\neq m,n} \(-(p_i +p_j)^2 \)^2 \(-(p_k+p_l)^2\)^2 \(-(p_m+p_n)^2\)^2\\
        p_F^{(6)}=& \sum_{i<j; k\neq i,j} \(-(p_i +p_j)^2\)^4 \(-(p_i+p_j+p_k)^2\)^2\\
        p_G^{(6)}=& \sum_{i<j; k\neq i,j} \(-(p_i +p_j)^2\)^3 \(-(p_i+p_j+p_k)^2\)^3\\
        p_H^{(6)}=& \sum_{i<j; k\neq i,j} \(-(p_i +p_j)^2\)^2 \(-(p_i+p_j+p_k)^2\)^4\\
        p_I^{(6)}=& \frac{1}{2}\sum_{i<j; k\neq i,j;l \neq i,j,k} \(-(p_i +p_j)^2\)^2 \(-(p_i+p_j+p_k)^2\)^2 \(-(p_i+p_j+p_k+p_l)^2\)^2\\
        p_J^{(6)}=& \frac{1}{2}\sum_{i<j<k} \(-(p_i+p_j+p_k)^2\)^6
    \end{array}
    $$
    \vfill
    \newpage

\section{Wilson Coefficients from UV Ansatz}\label{appendix coefficients uv}
    Here we list all Wilson coefficients that arise when expanding the heavy-exchange contributions in the five- and six-$\phi$ amplitudes using the basis of polynomials specified in appendix \ref{appendix explicit polynomials}. We make use of the shifted quartic coupling $\widetilde{\lambda}_a$ defined in \eqref{eq def shifted lambda}.
    
    \subsection{Five-Particle Scattering}
    
    \begin{flalign*}
    \begin{aligned} 
    \beta^{(0)}=&10 \sum_a \frac{g_a \lambda_a}{m_a^2} - 15\sum_{ab} \frac{g_a g_{ab} g_b}{m_a^2 m_b^2}\\
    \beta^{(2)}=& \sum_a \frac{g_a \lambda_a}{m_a^6} - \frac{1}{2}\sum_{ab} \frac{g_a g_{ab}g_b}{m_a^6 m_b^6} m_a^2 (6m_a^2+ m_b^2)\\
    \beta^{(3)}=& \sum_a\frac{g_a \lambda_a}{m_a^8} - \sum_{ab} \frac{g_a g_{ab}g_b}{m_a^8 m_b^8}m_a^4 (3m_a^2+ m_b^2)=\sum_a \frac{g_a \widetilde{\lambda}_a}{m_a^8}\\
    \beta_A^{(4)}=&\sum_a \frac{g_a \lambda_a}{m_a^{10}} - \sum_{ab} \frac{g_a g_{ab}g_b}{m_a^{10} m_b^{10}} m_a^6 (3m_a^2+m_b^2)=\sum_a \frac{g_a \widetilde{\lambda}_a}{m_a^{10}}\\
    \beta_B^{(4)}=& -\sum_{ab} \frac{g_a g_{ab}g_b}{m_a^6 m_b^6}
    \end{aligned}&&
    \end{flalign*}
    
    \subsection{Six-Particle Scattering}
    \begin{equation*}
    \begin{aligned}
         \gamma^{(0)}=& 10 \sum_a \frac{\lambda_a^2}{m_a^2} - 45\sum_{ab} \frac{g_a \lambda_{ab} g_b}{m_a^2 m_b^2} -60 \sum_{ab} \frac{g_a g_{ab} \lambda_b}{m_a^2 m_b^2} + 90 \sum_{abc} \frac{g_a g_{ab} g_{bc} g_c}{m_a^2 m_b^2 m_c^2} + 15 \sum_{abc} \frac{g_a g_b g_c g_{abc}}{m_a^2 m_b^2 m_c^2}\\
        \gamma^{(2)}=& \sum_a \frac{\lambda_a^2}{m_a^6} - \frac{1}{2} \sum_{ab} \frac{g_a \lambda_{ab} g_b}{m_a^6 m_b^6}m_a^2(12 m_a^2 +m_b^2) - 2 \sum_{ab} \frac{g_a g_{ab} \lambda_b}{m_a^6 m_b^6} (3 m_a^4 +m_a^2 m_b^2 +2 m_b^4)\\
        &+\sum_{abc} \frac{g_a g_{ab} g_{bc} g_c}{m_a^6 m_b^6 m_c^6} (m_a^2 m_c^2 (3m_a^2+m_b^2)(3m_c^2+m_b^2)+12 m_a^4 m_b^4) + \frac{1}{2}\sum_{abc} \frac{g_a g_b g_c g_{abc}}{m_a^6 m_b^6 m_c^6}m_a^4 m_b^2 (6m_b^2+ m_c^2)\\
        \gamma_A^{(3)}=& - \sum_{ab} \frac{g_a \lambda_{ab} g_b}{m_a^8 m_b^8} m_a^4\(6m_a^2+m_b^2 \) 
        -2 \sum_{ab} \frac{g_a g_{ab} \lambda_b}{m_a^8 m_b^4} \(m_a^2+2m_b^2\) \\
        &+ 2\sum_{abc} \frac{g_a g_{ab} g_{bc} g_c}{m_a^8 m_b^4 m_c^8} m_a^4 \(6 m_a^2 m_b^2+3 m_a^2 m_c^2+m_b^2 m_c^2\) + \frac{1}{3} \sum_{abc} \frac{g_a g_b g_c g_{abc}}{m_a^8 m_b^8 m_c^8} m_a^4 m_b^4 (9 m_a^2 m_b^2+3 m_a^2 m_c^2 + m_c^4) \\
        \gamma_B^{(3)}=&\sum_a \frac{\lambda_a^2}{m_a^8} - 2\sum_{ab} \frac{g_a g_{ab} \lambda_b}{m_a^4 m_b^8} (3m_a^2 + m_b^2) + \sum_{abc} \frac{g_a g_{ab} g_{bc} g_c}{m_a^4 m_b^8 m_c^4} (3m_a^2+m_b^2)(3m_c^2+m_b^2) \\
        &- \frac{1}{3} \sum_{abc} \frac{g_a g_b g_c g_{abc}}{m_a^4 m_b^4 m_c^4} = \sum_a \frac{\widetilde{\lambda}_a^2}{m_a^8}-\frac{1}{3} \sum_{abc} \frac{g_a g_b g_c g_{abc}}{m_a^4 m_b^4 m_c^4}\\    
    \end{aligned}
    \end{equation*}
    \vfill
\newpage
    \begin{equation*}
    \begin{aligned}
    \gamma_A^{(4)}=& -\sum_{ab}\frac{g_a \lambda_{ab} g_b}{m_a^{10}     m_b^{10}} m_a^6(6m_a^2+m_b^2) 
            - 2\sum_{ab} \frac{g_a g_{ab} \lambda_b}{m_a^{10} m_b^4}(m_a^2+ 2 m_b^2) \\
            &+ 2 \sum_{abc} \frac{g_a g_{ab} g_{bc} g_c}{m_a^{10} m_b^4 m_c^{10}} m_a^6 (6 m_a^2 m_b^2+ 3m_a^2 m_c^2 + m_b^2 m_c^2)\\
            &+ \frac{1}{2} \sum_{abc}\frac{g_a g_b g_c g_{abc}}{m_a^{10} m_b^{10} m_c^{10}} m_a^6 m_b^6 (6 m_a^2 m_b^2 +2 m_a^2 m_c^2 + m_c^4)  \\
    \gamma_B^{(4)}=& -\sum_{ab} \frac{g_a \lambda_{ab} g_b}{m_a^6 m_b^6} 
            + 2 \sum_{abc} \frac{g_a g_{ab} g_{bc} g_c}{m_a^6 m_b^2 m_c^6} 
            +\sum_{abc} \frac{g_a g_b g_c g_{abc}}{m_a^6 m_b^6 m_c^6} m_a^2(m_a^2+m_b^2)\\    
    \gamma_C^{(4)}=& - \sum_{ab} \frac{g_a g_{ab} \lambda_b}{m_a^6 m_b^6} 
            + \sum_{abc} \frac{g_a g_{ab} g_{bc} g_c}{m_a^6 m_b^6 m_c^6} m_a^2(3m_a^2 + m_b^2) 
            - \frac{1}{2} \sum_{abc} \frac{g_a g_b g_c g_{abc}}{m_a^6 m_b^6 m_c^6} m_a^2 m_b^2\\
            =& -\sum_{ab} \frac{g_a g_{ab} \widetilde{\lambda}_b}{m_a^6 m_b^6} - \frac{1}{2} \sum_{abc} \frac{g_a g_b g_c g_{abc}}{m_a^6 m_b^6 m_c^6} m_a^2 m_b^2\\
    \gamma_D^{(4)}=& \sum_a \frac{\lambda_a^2}{m_a^{10}}- 2\sum_{ab} \frac{g_a g_{ab}\lambda_b}{m_a^4 m_b^{10}}(3m_a^2 + m_b^2) +  \sum_{abc} \frac{g_a g_{ab} g_{bc} g_c}{m_a^4 m_b^{10} m_c^4} (3m_a^2 +m_b^2) (3 m_c^2 + m_b^2)= \sum_a \frac{\widetilde{\lambda}_a^2}{m_a^{10}}\\
    \gamma_A^{(5)}=&- \sum_{ab} \frac{g_a \lambda_{ab} g_b}{m_a^{12} m_b^{12}} m_a^8 (6 m_a^2   +m_b^2) -2 \sum_{ab} \frac{g_a g_{ab} \lambda_b}{m_a^{12} m_b^4} (m_a^2 +2m_b^2)\\
        &+2 \sum_{abc} \frac{g_a g_{ab}g_{bc} g_c}{m_a^{12}m_b^4 m_c^{12}} m_a^8 (6 m_a^2 m_b^2+3 m_a^2 m_c^2+m_b^2 m_c^2)\\
        &+\frac{1}{2} \sum_{abc}\frac{g_a g_b g_c g_{abc}}{m_a^{12} m_b^{12} m_c^{12}} m_a^8 m_b^8 (6 m_a^2 m_b^2 + 2 m_a^2 m_c^2 +m_c^4)\\
        \gamma_B^{(5)}=& -\sum_{ab} \frac{g_a \lambda_{ab}g_b}{m_a^8 m_b^8} m_a^2 
        + \sum_{abc} \frac{g_a g_{ab} g_{bc} g_c}{m_a^8 m_b^4 m_c^8}m_a^2 (2 m_b^2 + m_c^2)
        +\frac{1}{2} \sum_{abc} \frac{g_a g_b g_c g_{abc}}{m_a^8 m_b^8 m_c^8} m_a^4 m_b^2(2 m_a^2 +m_b^2)\\
        \gamma_C^{(5)}=& - \sum_{abc} \frac{g_a g_{ab} g_{bc} g_c}{m_a^6 m_b^4 m_c^6} + \sum_{abc} \frac{g_a g_b g_c g_{abc}}{m_a^6 m_b^6 m_c^6} m_a^2 \\
        \gamma_D^{(5)}=& -\sum_{ab} \frac{g_a g_{ab}\lambda_b}{m_a^8 m_b^6} 
        + \sum_{abc} \frac{g_a g_{ab} g_{bc} g_c}{m_a^8 m_b^6 m_c^8} m_a^4 (3m_a^2 +m_b^2) - \frac{1}{2}\sum_{abc}\frac{g_a g_b g_c g_{abc}}{m_a^8 m_b^8 m_c^8} m_a^4 m_b^4\\
        =& -\sum_{ab}\frac{g_a g_{ab} \widetilde{\lambda}_b}{m_a^8 m_b^6} - \frac{1}{2} \sum_{abc} \frac{g_a g_b g_c g_{abc}}{m_a^8 m_b^8 m_c^8} m_a^4 m_b^4\\
        \gamma_E^{(5)}=&\sum_a \frac{\lambda_a^2}{m_a^{12}}
        -2 \sum_{ab} \frac{g_a g_{ab} \lambda_b}{m_a^4 m_b^{12}}(3m_a^2 +m_b^2) 
        +\sum_{abc} \frac{g_a g_{ab} g_{bc} g_c}{m_a^4 m_b^{12} m_c^4} (3m_a^2+m_b^2) (3m_c^2+m_b^2)= \sum_a \frac{\widetilde{\lambda}_a^2}{m_a^{12}}\\
        \gamma_F^{(5)}=& - \sum_{ab} \frac{g_a g_{ab}\lambda_b}{m_a^6 m_b^8} 
        + \sum_{abc} \frac{g_a g_{ab} g_{bc} g_c}{m_a^6 m_b^8 m_c^6} m_a^2 (3m_a^2+m_b^2)=-\sum_{ab} \frac{g_a g_{ab} \widetilde{\lambda}_b}{m_a^6 m_b^8}\\
    \gamma_A^{(6)}=& -\sum_{ab} \frac{g_a \lambda_{ab} g_b}{m_a^{14} m_b^{14}} m_a^{10} (6 m_a^2+m_b^2) 
        -2 \sum_{ab} \frac{g_a g_{ab}\lambda_b}{m_a^{14}m_b^4}(m_a^2 + 2 m_b^2)\\
        &+ 2 \sum_{abc} \frac{g_a g_{ab} g_{bc} g_c}{m_a^{14}m_b^4 m_c^{14}} m_a^{10} (6 m_a^2 m_b^2 +3 m_a^2 m_c^2 +m_b^2 m_c^2)\\
        &+\frac{1}{2} \sum_{abc} \frac{g_a g_b g_c g_{abc}}{m_a^{14} m_b^{14} m_c^{14}} m_a^{10} m_b^{10} (6 m_a^2 m_b^2+2 m_a^2 m_c^2+m_c^4) \\
    \gamma_B^{(6)}=& -\sum_{ab} \frac{g_a \lambda_{ab} g_b}{m_a^{10} m_b^{10}} m_a^4 +\sum_{abc} \frac{g_a g_{ab} g_{bc} g_c}{m_a^{10} m_b^4 m_c^{10}} m_a^4 (2m_b^2+m_c^2) 
        +\frac{1}{2} \sum_{abc} \frac{g_a g_b g_c g_{abc}}{m_a^{10} m_b^{10} m_c^{10}} m_a^6 m_b^4 (2m_a^2 + m_b^2) \\
    \end{aligned}
    \end{equation*}

    \begin{equation*}
    \begin{aligned}
        \gamma_C^{(6)}=& - \sum_{ab} \frac{g_a \lambda_{ab} g_b}{m_a^8 m_b^8} +2 \sum_{abc} \frac{g_a g_{ab} g_{bc} g_c}{m_a^8 m_b^4 m_c^8} (m_a^2+m_b^2) 
        +\sum_{abc} \frac{g_a g_b g_c g_{abc}}{m_a^8 m_b^8 m_c^8} m_a^6\\
        \gamma_D^{(6)}=&-\sum_{abc} \frac{g_a g_{ab} g_{bc} g_c}{m_a^8 m_b^4 m_c^8} m_a^2 + \sum_{abc} \frac{g_a g_b g_c g_{abc}}{m_a^8 m_b^8 m_c^8} m_a^4 m_b^2 \\
        \gamma_E^{(6)}=& \sum_{abc} \frac{g_a g_b g_c g_{abc}}{m_a^6 m_b^6 m_c^6}\\
        \gamma_F^{(6)}=& -\sum_{ab} \frac{g_a g_{ab} \lambda_b}{m_a^{10} m_b^6}
        + \sum_{abc}\frac{g_a g_{ab} g_{bc} g_c}{m_a^{10} m_b^6 m_c^{10}} m_a^6 (3m_a^2 +m_b^2) 
        -\frac{1}{2}\sum_{abc} \frac{g_a g_b g_c g_{abc}}{m_a^{10} m_b^{10} m_c^{10}} m_a^6 m_b^6\\
        =& -\sum_{ab} \frac{g_a g_{ab} \widetilde{\lambda}_b}{m_a^{10}m_b^6}-\frac{1}{2}\sum_{abc} \frac{g_a g_b g_c g_{abc}}{m_a^{10}m_b^{10}m_c^{10}}m_a^6 m_b^6\\
        \gamma_G^{(6)}=&-\sum_{ab} \frac{g_a g_{ab} \lambda_b}{m_a^8 m_b^8} +\sum_{abc} \frac{g_a g_{ab} g_{bc} g_c}{m_a^8 m_b^8 m_c^8} m_a^4 (3m_a^2+m_b^2)= -\sum_{ab} \frac{g_a g_{ab}\widetilde{\lambda}_b}{m_a^8 m_b^8}\\
        \gamma_H^{(6)}=&- \sum_{ab}  \frac{g_a g_{ab} \lambda_b}{m_a^6 m_b^{10}}+ \sum_{abc} \frac{g_a g_{ab} g_{bc} g_c}{m_a^6 m_b^{10} m_c^6} m_a^2 (3m_a^2 +m_b^2)=-\sum_{ab} \frac{g_a g_{ab}\widetilde{\lambda}_b}{m_a^6 m_b^{10}}   \\
        \gamma_I^{(6)}=& \sum_{abc} \frac{g_a g_{ab} g_{bc} g_c}{m_a^6 m_b^6 m_c^6}    \\
        \gamma_J^{(6)}=& \sum_a \frac{\lambda_a^2}{m_a^{14}} 
        -2\sum_{ab} \frac{g_a g_{ab}\lambda_b}{m_a^4 m_b^{14}}(3m_a^2+m_b^2)
        +\sum_{abc} \frac{g_a g_{ab} g_{bc} g_c}{m_a^4 m_b^{14} m_c^4} (3m_a^2 +m_b^2)(3m_c^2 +m_b^2)=\sum_a \frac{\widetilde{\lambda}_a^2}{m_a^{14}}\\
    \end{aligned}
    \end{equation*}
    
\section{Separating Light- and Heavy-Exchange Contributions}\label{appendix light exchanges}
One of the key ingredients in interpreting the stated inequalities as bounds on amplitude coefficients instead of channel residues stems from how we account for light-particle exchange. While we commented in great detail about the prescription used to extract analytic contributions, we want to proceed here by explaining how we established the stated independence to the bounded amplitude contributions.\\

After applying the prescription described in section \ref{sect light exchanges uv}, we obtain a new set of contributions to the analytic part to the scattering amplitude at $n$-th order in derivative expansion $\mathcal{A}^{(n)}$ 
\begin{equation}\label{gen amp appendix}
   \mathcal{A}^{(n)} = \sum_i\beta_i^{(n)} q_i^{(n)} + \sum_I\beta_I^{(n)} q_I^{(n)},
\end{equation}
with $q_i^{(n)}$ being all light-exchange contributions and $q_I^{(n)}$ the independent contributions listed in the Appendix~\ref{appendix explicit polynomials}. However, the different coefficients $\beta_i^{(n)}$, $\beta_I^{(n)}$ can only be read from the amplitude if all polynomials in which $\mathcal{A}^{(n)}$ are expanded are linearly independent. If they are not, there must be a relation
\begin{equation}\label{independence statement appendix}
    \sum_i \lambda_i q_i^{(n)} +\sum_I \lambda_I q_I^{(n)}=0,
\end{equation}
for some real numbers $\lambda_i$, $\lambda_I$. Assuming, for example, $\lambda_a\neq 0$, we can solve this equation for $q_a^{(n)}$, which will remove one term from \eqref{gen amp appendix}, while shifting all other coefficients accordingly.
\begin{equation}
    \beta_i^{(n)}\rightarrow\beta_i^{(n)}-\frac{\lambda_i}{\lambda_a}\beta_a^{(n)} \qquad \beta_I^{(n)}\rightarrow\beta_I^{(n)}-\frac{\lambda_I}{\lambda_a}\beta_a^{(n)}
\end{equation}
In order to check whether there is a set of numbers that fulfil \eqref{independence statement appendix}, we set up a system of linear equations. For this, note that \eqref{independence statement appendix} is equivalent to the coefficients of each monomial in $q_i^{(n)}$, $q_I^{(n)}$ being related by the same set of numbers $\lambda_i$, $\lambda_I$. 

To make things as explicit as possible, the analytic contribution to five-point scattering at fourth order in $\frac{\Box}{m_a^2}$ is given by  
\begin{eqnarray}
     \mathcal{A}_{5\phi}^{(4)} =&& \beta_l^{(4)} q_l^{(4)} + \beta_A^{(4)} q_A^{(4)}+ \beta_B^{(4)} q_B^{(4)},\qquad \text{with}\\\nn
     q_l^{(4)} =&&19 s_1^4 - 28 s_1^3 s_2 + 42 s_1^2 s_2^2 - 28 s_1 s_2^3 + 19 s_2^4 + 14 s_1^3 t_{23} - 42 s_1^2 s_2 t_{23} + 42 s_1 s_2^2 t_{23} - 38 s_2^3 t_{23} \\\nn
     && + 66 s_1^2 t_{23}^2 - 52 s_1 s_2 t_{23}^2 + 62 s_2^2 t_{23}^2 + 19 s_1 t_{23}^3 - 43 s_2 t_{23}^3 + 24 t_{23}^4 + 34 s_1^3 t_{24} + 34 s_2^3 t_{24} \\ \nn
     &&- 45 s_1^2 t_{23} t_{24} + 10 s_1 s_2 t_{23} t_{24} - 82 s_2^2 t_{23} t_{24} + 35 s_1 t_{23}^2 t_{24} + 77 s_2 t_{23}^2 t_{24} - 48 t_{23}^3 t_{24} + 51 s_1^2 t_{24}^2\\ \nn
     &&+ 51 s_2^2 t_{24}^2 - 40 s_1 t_{23} t_{24}^2 - 77 s_2 t_{23} t_{24}^2 + 67 t_{23}^2 t_{24}^2 + 34 s_1 t_{24}^3 + 34 s_2 t_{24}^3 - 43 t_{23} t_{24}^3 + 19 t_{24}^4 \\ \nn
     &&- 43 s_1^3 t_{34} + 62 s_1^2 s_2 t_{34} - 72 s_1 s_2^2 t_{34} + 34 s_2^3 t_{34} + 30 s_1^2 t_{23} t_{34} + 30 s_1 s_2 t_{23} t_{34} - 20 s_2^2 t_{23} t_{34} \\ \nn
     &&- 60 s_1 t_{23}^2 t_{34} + 25 s_2 t_{23}^2 t_{34} + 29 t_{23}^3 t_{34} - 72 s_1^2 t_{24} t_{34} + 90 s_1 t_{23} t_{24} t_{34} - 10 s_2 t_{23} t_{24} t_{34} \\ \nn
     &&- 62 t_{23}^2 t_{24} t_{34} - 72 s_1 t_{24}^2 t_{34} + 62 t_{23} t_{24}^2 t_{34} - 28 t_{24}^3 t_{34} + 62 s_1^2 t_{34}^2 - 72 s_1 s_2 t_{34}^2 + 51 s_2^2 t_{34}^2 \\ \nn
     && - 30 s_1 t_{23} t_{34}^2 - 15 s_2 t_{23} t_{34}^2 + 66 t_{23}^2 t_{34}^2 + 62 s_1 t_{24} t_{34}^2 - 72 t_{23} t_{24} t_{34}^2 + 42 t_{24}^2 t_{34}^2 - 43 s_1 t_{34}^3\\ \nn
     &&+ 34 s_2 t_{34}^3 + 34 t_{23} t_{34}^3 - 28 t_{24} t_{34}^3 + 19 t_{34}^4,
\end{eqnarray}
and $q_A^{(4)}$, $q_B^{(4)}$ given in appendix \ref{appendix explicit polynomials}. To check their independence, we set up a system of linear equations 
\begin{eqnarray}
    &&\lambda_l a_l +\lambda_A a_A + \lambda_B a_B=0 \\ \nn
    &&\lambda_l b_l +\lambda_A b_A + \lambda_B b_B=0 \\ \nn
    &&\lambda_l c_l +\lambda_A c_A + \lambda_B c_B=0 \,,
\end{eqnarray}
with $a_l/a_A/a_B$ being the coefficients of some monomial in $q_l^{(4)}/q_A^{(4)}/q_B^{(4)}$ and similarly for the different $b_i$ and $c_i$. Taking into account, for example, the three monomials $s_1^4$, $t_{23} t_{24} t_{34}^2$ and $s_1 t_{23} t_{24} t_{34}$, this system of equations becomes
\begin{eqnarray}
    &&  19 \lambda_l+ 2 \lambda_A + 2 \lambda_B  = 0\\\nn
    &&- 72 \lambda_l -6 \lambda_A - 6 \lambda_B  = 0\\\nn
    &&  90 \lambda_l +4 \lambda_B = 0
\end{eqnarray}
which only admits the trivial solution. We conclude that the three considered polynomials are indeed independent and that we can therefore unambiguously isolate the coefficients $\beta_A^{(4)}$ and $\beta_B^{(4)}$ that enter into the positive matrices \eqref{bound matrix} and \eqref{bound matrix2}.

At fourth order in $\frac{\Box}{m_a^2}$, we start out with a total of eight polynomials that capture light exchanges via the prescription outlined in section \ref{sect light exchanges uv}. We refrain from listing them here given their size. After doing so, we observe one relation involving terms in the light sector, as well as a second identity that relates some light polynomials to $p_A^{(4)}$. The minimal list of generators therefore involves a total of ten polynomials. These ten polynomials will later be matched to the EFT. At fifth order in Mandelstams we start out with a total of ten light polynomials together with the five heavy-exchange contributions that remain after imposing the Gram constraint \eqref{gram at fifth order} to solve for $p_F^{(5)}$ in terms of the others. The independence check derives two relations among them, one relating $p_A^{(5)}$ to some light contributions and another equation that only involves light polynomials. This reduces the total number of generators after imposing the Gram constraint to 13. At sixth order in the expansion we again obtained ten light-exchange polynomials using the outlined prescription. As was the case before, two constraints have been found, one of which includes the heavy polynomial $p_A^{(6)}$. The total rank of UV generators therefore equals 18.\\
The conclusion from these files is that mixing does occur, particularly between $p_A^{(4)}$ , $p_A^{(5)}$ and $p_A^{(6)}$ listed in appendix \ref{appendix explicit polynomials} and some quartic/quintic/sextic polynomial arising from light exchanges. Crucially, none of the obtained relations involve any of the polynomials $p_D^{(4)}$, $p_E^{(5)}$, $p_H^{(6)}$, $p_I^{(6)}$, $p_J^{(6)}$ whose expansion coefficients appear in \eqref{bound matrix} and \eqref{bound matrix2}. Their associated coefficients can therefore all be unambiguously determined from the scattering amplitude when expanding in the corresponding basis.

\section{Basis of Derivative Interactions for Single-Scalar EFT}\label{appendix eft interactions}
    We list here a full basis of all relevant derivative interactions to evaluate the six-$\phi$ scattering amplitude in a single-scalar EFT up to ${\cal O}(\Lambda^{-14})$. We use this basis in order to calculate the contributions arising from the light-exchange diagrams stated in section \ref{sect eval eft}. We then extract their analytic contribution via the same prescription used in the UV, which is detailed in section \ref{sect light exchanges uv} . We then check the linear independence of these light-exchange terms together with EFT contact terms at any given order. \\
    We want to stress that by working in $D=4$, the nontrivial Gram constraint induces a relation among the otherwise ${\cal O}(\Lambda^{-12})$ interactions among six fields. This is the reason why we only have five independent terms at this order, instead of the six that would arise for $D\geq 5$.\\\\
    We denote $(\partial^{(n)} \phi)^2 \equiv \partial_{\mu_1..\mu_n}\phi \partial^{\mu_1...\mu_n}\phi$, $(\Delta\phi)_{\alpha \beta} \equiv\partial_{\alpha \beta} \phi$,
    $(\widetilde{\Delta}\phi)_{\alpha \beta}\equiv \partial_{\alpha \beta \gamma}\phi \partial^\gamma \phi$,\linebreak $(\widetilde{\widetilde{\Delta}}\phi)_{\alpha \beta}\equiv \partial_{\alpha \beta \gamma \delta}\phi \partial^{\gamma \delta}\phi$.    
    Using these operators, we write $\text{Tr}(\Delta^2 \phi)^3\equiv \partial_{\alpha\beta\gamma \delta}\phi \partial_{\gamma \delta \epsilon \varphi} \phi \partial_{\epsilon\varphi \alpha \beta}\phi$ and similarly for $\text{Tr}(\Delta^2 \phi)^4$.
    \\
\subsection{Four Fields}
    \begin{flalign*}
    \begin{array}{ll}
        {\cal O}(\Lambda^{-4}): & \frac{a^{(2)}}{8} (\partial \phi)^2(\partial \phi)^2\\
        {\cal O}(\Lambda^{-6}): & \frac{a^{(3)}}{4} (\partial \phi)^2(\partial^{(2)}\phi)^2\\
        {\cal O}(\Lambda^{-8}): & \frac{a^{(4)}}{8} (\partial^{(2)}\phi)^2(\partial^{(2)}\phi)^2\\
        {\cal O}(\Lambda^{-10}): & \frac{a^{(5)}}{4} (\partial^{(2)} \phi)^2(\partial^{(3)}\phi)^2\\
        {\cal O}(\Lambda^{-12}): & \frac{a_A^{(6)}}{8} (\partial^{(3)} \phi)^2(\partial^{(3)}\phi)^2 \qquad \frac{a_B^{(6)}}{3!} \phi \text{Tr}(\Delta^2\phi)^3\\
        {\cal O}(\Lambda^{-14}): & \frac{a^{(7)}}{4} (\partial^{(3)} \phi)^2(\partial^{(4)}\phi)^2\\
        {\cal O}(\Lambda^{-16}): & \frac{a_A^{(8)}}{8} (\partial^{(4)} \phi)^2(\partial^{(4)}\phi)^2 \qquad \frac{a_B^{(8)}}{4!} \text{Tr}(\Delta^2\phi)^4
    \end{array}&&
    \end{flalign*}

\subsection{Five Fields}
    \begin{flalign*}
    \begin{array}{ll}
        {\cal O}(\Lambda^{-5}): & \frac{b^{(2)}}{8}\phi (\partial\phi)^2 (\partial \phi)^2\\
        {\cal O}(\Lambda^{-7}): & \frac{b^{(3)}}{4}\phi (\partial\phi)^2 (\partial^{(2)} \phi)^2\\
        {\cal O}(\Lambda^{-9}): &\frac{b_A^{(4)}}{4}\phi (\partial\phi)^2 (\partial^{(3)} \phi)^2 \qquad \frac{b_B^{(4)}}{8}\phi (\partial^{(2)} \phi)^2 (\partial^{(2)} \phi)^2\\
        {\cal O}(\Lambda^{-11}): & \frac{b_A^{(5)}}{4}\phi (\partial\phi)^2 (\partial^{(4)} \phi)^2 \qquad \frac{b_B^{(5)}}{4}\phi (\partial^{(2)} \phi)^2 (\partial^{(3)} \phi)^2\\
        {\cal O}(\Lambda^{-13}): & \frac{b_A^{(6)}}{4}\phi (\partial\phi)^2 (\partial^{(5)} \phi)^2 \qquad \frac{b_B^{(6)}}{4}\phi (\partial^{(2)} \phi)^2 (\partial^{(4)} \phi)^2 \qquad \frac{b_C^{(6)}}{8}\phi (\partial^{(3)} \phi)^2 (\partial^{(3)} \phi)^2 \qquad
        \frac{b_D^{(6)}}{2 \, 3!}\phi^2 \text{Tr}(\Delta^2 \phi)^3 \\
        & \frac{b_E^{(6)}}{2} \phi \text{Tr} (\widetilde{\widetilde{\Delta}}\phi )^2\\
        {\cal O}(\Lambda^{-15}): & \frac{b_A^{(7)}}{4}\phi (\partial\phi)^2 (\partial^{(6)} \phi)^2 \qquad \frac{b_B^{(7)}}{4}\phi (\partial^{(2)} \phi)^2 (\partial^{(5)} \phi)^2 \qquad \frac{b_C^{(7)}}{4} \phi (\partial^{(3)} \phi)^2 (\partial^{(4)} \phi)^2 \qquad
        \frac{b_D^{(7)}}{2 \, 3!}(\partial\phi)^2 \text{Tr}(\Delta^2 \phi)^3\\
    \end{array}&&
    \end{flalign*}
    
    \subsection{Six Fields}
    \begin{flalign*}
    \begin{array}{ll}
        {\cal O}(\Lambda^{-6}):& \frac{c^{(2)}}{16} \phi^2 (\partial \phi)^2 (\partial \phi)^2\\
        {\cal O}(\Lambda^{-8}):& \frac{c_A^{(3)}}{8} \phi^2 (\partial \phi)^2 (\partial^{(2)} \phi)^2 \qquad \frac{c_B^{(3)}}{2 \, 3! \, 3!} \phi^3 \Box^3 \phi^3\\
        {\cal O}(\Lambda^{-10}): &\frac{c_A^{(4)}}{8} \phi^2 (\partial\phi)^2 (\partial^{(3)} \phi)^2 \qquad \frac{c_B^{(4)}}{16} \phi^2(\partial^{(2)} \phi)^2(\partial^{(2)} \phi)^2 \qquad \frac{c_C^{(4)}}{2\, 3!} \phi^3 \Box^2\( \phi (\partial^{(2)}\phi)^2\)\qquad \frac{c_D^{(4)}}{2 \, 3! \, 3!} \phi^3 \Box^4 \phi^3\\
        {\cal O}(\Lambda^{-12}):& \frac{c_A^{(5)}}{8} \phi^2 (\partial\phi)^2 (\partial^{(4)} \phi)^2 \qquad \frac{c_B^{(5)}}{8} \phi^2(\partial^{(2)} \phi)^2(\partial^{(3)} \phi)^2 \qquad \frac{c_C^{(5)}}{16} (\partial \phi)^2(\partial^{(2)}\phi)^2(\partial^{(2)}\phi)^2 \\ 
        & \frac{c_D^{(5)}}{2 \, 3!} \phi^3 \Box^2\( \phi (\partial^{(3)}\phi)^2\) \qquad \frac{c_E^{(5)}}{2 \, 3! \, 3!} \phi^3 \Box^5 \phi^3\\
        {\cal O}(\Lambda^{-14}):& \frac{c_A^{(6)}}{8} \phi^2 (\partial\phi)^2 (\partial^{(5)} \phi)^2 \qquad \frac{c_B^{(6)}}{8} \phi^2(\partial^{(2)} \phi)^2(\partial^{(4)} \phi)^2 \qquad \frac{c_C^{(6)}}{16} \phi^2(\partial^{(3)} \phi)^2(\partial^{(3)} \phi)^2 \\
        & \frac{c_D^{(6)}}{8} (\partial \phi)^2 (\partial^{(2)}\phi)^2 (\partial^{(3)}\phi)^2 \qquad \frac{c_E^{(6)}}{8\, 3!} (\partial^{(2)}\phi)^2 (\partial^{(2)}\phi)^2 (\partial^{(2)}\phi)^2 \qquad \frac{c_F^{(6)}}{2 \, 3!} \phi^3 \Box^2\( \phi (\partial^{(4)}\phi)^2\) \\
        & \frac{c_G^{(6)}}{2 \, 3!} \phi^3 \Box^3\( \phi (\partial^{(3)}\phi)^2\) \qquad \frac{c_H^{(6)}}{2 \, 3!} \phi^3 \Box^4\( \phi (\partial^{(2)}\phi)^2\) \qquad \frac{c_I^{(6)}}{8} \phi (\partial^{(2)}\phi)^2 \Box^2\( \phi (\partial^{(2)}\phi)^2\) \\
        & \frac{c_J^{(6)}}{{2 \, 3! \, 3!}} \phi^3 \Box^6 \phi^3 \qquad \frac{c_K^{(6)}}{3! \, 3!} \phi^3 \text{Tr}(\Delta^2\phi)^3 \qquad \frac{c_L^{(6)}}{3!}\text{Tr}(\widetilde{\Delta} \phi)^3 \qquad  \frac{c_M^{(6)}}{4} \phi^2 \text{Tr} (\widetilde{\widetilde{\Delta}}\phi)^2
    \end{array}&&
    \end{flalign*}


\bibliographystyle{JHEP}

\bibliography{biblio.bib}

\providecommand{\href}[2]{#2}\begingroup\raggedright\begin{thebibliography}{10}

\bibitem{Yndurain:1969qm}
F.J.~Yndurain, \emph{{Constraints on pi pi partial waves from positivity and analyticity}}, \href{https://doi.org/10.1007/BF02824574}{\emph{Nuovo Cim. A} {\bfseries 64} (1969) 225}.

\bibitem{Common:1970ck}
A.K.~Common and F.J.~Yndurain, \emph{{Constraints on the derivatives of the pi pi scattering amplitude from positivity}}, \href{https://doi.org/10.1007/BF01646093}{\emph{Commun. Math. Phys.} {\bfseries 18} (1970) 171}.

\bibitem{Yndurain:1972ix}
F.J.~Yndurain, \emph{{Rigorous constraints, bounds, and relations for scattering amplitudes}}, \href{https://doi.org/10.1103/RevModPhys.44.645}{\emph{Rev. Mod. Phys.} {\bfseries 44} (1972) 645}.

\bibitem{Pham:1985cr}
T.N.~Pham and T.N.~Truong, \emph{{Evaluation of the Derivative Quartic Terms of the Meson Chiral Lagrangian From Forward Dispersion Relation}}, \href{https://doi.org/10.1103/PhysRevD.31.3027}{\emph{Phys. Rev. D} {\bfseries 31} (1985) 3027}.

\bibitem{Pennington:1994kc}
M.R.~Pennington and J.~Portoles, \emph{{The Chiral Lagrangian parameters, l1, l2, are determined by the rho resonance}}, \href{https://doi.org/10.1016/0370-2693(94)01551-M}{\emph{Phys. Lett. B} {\bfseries 344} (1995) 399} [\href{https://arxiv.org/abs/hep-ph/9409426}{{\ttfamily hep-ph/9409426}}].

\bibitem{Ananthanarayan:1994hf}
B.~Ananthanarayan, D.~Toublan and G.~Wanders, \emph{{Consistency of the chiral pion pion scattering amplitudes with axiomatic constraints}}, \href{https://doi.org/10.1103/PhysRevD.51.1093}{\emph{Phys. Rev. D} {\bfseries 51} (1995) 1093} [\href{https://arxiv.org/abs/hep-ph/9410302}{{\ttfamily hep-ph/9410302}}].

\bibitem{Dita:1998mh}
P.~Dita, \emph{{Positivity constraints on chiral perturbation theory pion pion scattering amplitudes}}, \href{https://doi.org/10.1103/PhysRevD.59.094007}{\emph{Phys. Rev. D} {\bfseries 59} (1999) 094007} [\href{https://arxiv.org/abs/hep-ph/9809568}{{\ttfamily hep-ph/9809568}}].

\bibitem{adams2006causality}
A.~Adams, N.~Arkani-Hamed, S.~Dubovsky, A.~Nicolis and R.~Rattazzi, \emph{{Causality, analyticity and an IR obstruction to UV completion}}, \href{https://doi.org/10.1088/1126-6708/2006/10/014}{\emph{JHEP} {\bfseries 10} (2006) 014} [\href{https://arxiv.org/abs/hep-th/0602178}{{\ttfamily hep-th/0602178}}].

\bibitem{deRham:2017avq}
C.~de~Rham, S.~Melville, A.J.~Tolley and S.-Y.~Zhou, \emph{{Positivity bounds for scalar field theories}}, \href{https://doi.org/10.1103/PhysRevD.96.081702}{\emph{Phys. Rev. D} {\bfseries 96} (2017) 081702} [\href{https://arxiv.org/abs/1702.06134}{{\ttfamily 1702.06134}}].

\bibitem{deRham:2017zjm}
C.~de~Rham, S.~Melville, A.J.~Tolley and S.-Y.~Zhou, \emph{{UV complete me: Positivity Bounds for Particles with Spin}}, \href{https://doi.org/10.1007/JHEP03(2018)011}{\emph{JHEP} {\bfseries 03} (2018) 011} [\href{https://arxiv.org/abs/1706.02712}{{\ttfamily 1706.02712}}].

\bibitem{Remmen:2020vts}
G.N.~Remmen and N.L.~Rodd, \emph{{Flavor Constraints from Unitarity and Analyticity}}, \href{https://doi.org/10.1103/PhysRevLett.127.149901}{\emph{Phys. Rev. Lett.} {\bfseries 125} (2020) 081601} [\href{https://arxiv.org/abs/2004.02885}{{\ttfamily 2004.02885}}].

\bibitem{Zhang:2020jyn}
C.~Zhang and S.-Y.~Zhou, \emph{{Convex Geometry Perspective on the (Standard Model) Effective Field Theory Space}}, \href{https://doi.org/10.1103/PhysRevLett.125.201601}{\emph{Phys. Rev. Lett.} {\bfseries 125} (2020) 201601} [\href{https://arxiv.org/abs/2005.03047}{{\ttfamily 2005.03047}}].

\bibitem{Bellazzini:2020cot}
B.~Bellazzini, J.~Elias~Mir{\'o}, R.~Rattazzi, M.~Riembau and F.~Riva, \emph{{Positive moments for scattering amplitudes}}, \href{https://doi.org/10.1103/PhysRevD.104.036006}{\emph{Phys. Rev. D} {\bfseries 104} (2021) 036006} [\href{https://arxiv.org/abs/2011.00037}{{\ttfamily 2011.00037}}].

\bibitem{Tolley:2020gtv}
A.J.~Tolley, Z.-Y.~Wang and S.-Y.~Zhou, \emph{{New positivity bounds from full crossing symmetry}}, \href{https://doi.org/10.1007/JHEP05(2021)255}{\emph{JHEP} {\bfseries 05} (2021) 255} [\href{https://arxiv.org/abs/2011.02400}{{\ttfamily 2011.02400}}].

\bibitem{Caron-Huot:2020cmc}
S.~Caron-Huot and V.~Van~Duong, \emph{{Extremal Effective Field Theories}}, \href{https://doi.org/10.1007/JHEP05(2021)280}{\emph{JHEP} {\bfseries 05} (2021) 280} [\href{https://arxiv.org/abs/2011.02957}{{\ttfamily 2011.02957}}].

\bibitem{Sinha:2020win}
A.~Sinha and A.~Zahed, \emph{{Crossing Symmetric Dispersion Relations in Quantum Field Theories}}, \href{https://doi.org/10.1103/PhysRevLett.126.181601}{\emph{Phys. Rev. Lett.} {\bfseries 126} (2021) 181601} [\href{https://arxiv.org/abs/2012.04877}{{\ttfamily 2012.04877}}].

\bibitem{Arkani-Hamed:2020blm}
N.~Arkani-Hamed, T.-C.~Huang and Y.-t.~Huang, \emph{{The EFT-Hedron}}, \href{https://doi.org/10.1007/JHEP05(2021)259}{\emph{JHEP} {\bfseries 05} (2021) 259} [\href{https://arxiv.org/abs/2012.15849}{{\ttfamily 2012.15849}}].

\bibitem{Li:2021lpe}
X.~Li, H.~Xu, C.~Yang, C.~Zhang and S.-Y.~Zhou, \emph{{Positivity in Multifield Effective Field Theories}}, \href{https://doi.org/10.1103/PhysRevLett.127.121601}{\emph{Phys. Rev. Lett.} {\bfseries 127} (2021) 121601} [\href{https://arxiv.org/abs/2101.01191}{{\ttfamily 2101.01191}}].

\bibitem{Chiang:2021ziz}
L.-Y.~Chiang, Y.-t.~Huang, W.~Li, L.~Rodina and H.-C.~Weng, \emph{{Into the EFThedron and UV constraints from IR consistency}}, \href{https://doi.org/10.1007/JHEP03(2022)063}{\emph{JHEP} {\bfseries 03} (2022) 063} [\href{https://arxiv.org/abs/2105.02862}{{\ttfamily 2105.02862}}].

\bibitem{Alberte:2021dnj}
L.~Alberte, C.~de~Rham, S.~Jaitly and A.J.~Tolley, \emph{{Reverse Bootstrapping: IR Lessons for UV Physics}}, \href{https://doi.org/10.1103/PhysRevLett.128.051602}{\emph{Phys. Rev. Lett.} {\bfseries 128} (2022) 051602} [\href{https://arxiv.org/abs/2111.09226}{{\ttfamily 2111.09226}}].

\bibitem{deRham:2022hpx}
C.~de~Rham, S.~Kundu, M.~Reece, A.J.~Tolley and S.-Y.~Zhou, \emph{{Snowmass White Paper: UV Constraints on IR Physics}},  in \emph{{Snowmass 2021}}, 3, 2022 [\href{https://arxiv.org/abs/2203.06805}{{\ttfamily 2203.06805}}].

\bibitem{deRham:2022sdl}
C.~de~Rham, L.~Engelbrecht, L.~Heisenberg and A.~L{\"u}scher, \emph{{Positivity bounds in vector theories}}, \href{https://doi.org/10.1007/JHEP12(2022)086}{\emph{JHEP} {\bfseries 12} (2022) 086} [\href{https://arxiv.org/abs/2208.12631}{{\ttfamily 2208.12631}}].

\bibitem{Hong:2023zgm}
D.-Y.~Hong, Z.-H.~Wang and S.-Y.~Zhou, \emph{{Causality bounds on scalar-tensor EFTs}}, \href{https://doi.org/10.1007/JHEP10(2023)135}{\emph{JHEP} {\bfseries 10} (2023) 135} [\href{https://arxiv.org/abs/2304.01259}{{\ttfamily 2304.01259}}].

\bibitem{DeAngelis:2023bmd}
S.~De~Angelis and G.~Durieux, \emph{{EFT matching from analyticity and unitarity}}, \href{https://doi.org/10.21468/SciPostPhys.16.3.071}{\emph{SciPost Phys.} {\bfseries 16} (2024) 071} [\href{https://arxiv.org/abs/2308.00035}{{\ttfamily 2308.00035}}].

\bibitem{Wan:2024eto}
S.-L.~Wan and S.-Y.~Zhou, \emph{{Matrix moment approach to positivity bounds and UV reconstruction from IR}}, \href{https://doi.org/10.1007/JHEP02(2025)168}{\emph{JHEP} {\bfseries 02} (2025) 168} [\href{https://arxiv.org/abs/2411.11964}{{\ttfamily 2411.11964}}].

\bibitem{Cheung:2025krg}
C.~Cheung and G.N.~Remmen, \emph{{Multipositivity bounds for scattering amplitudes}}, \href{https://doi.org/10.1103/wt4x-2149}{\emph{Phys. Rev. D} {\bfseries 112} (2025) 016017} [\href{https://arxiv.org/abs/2505.05553}{{\ttfamily 2505.05553}}].

\bibitem{Paulos:2016fap}
M.F.~Paulos, J.~Penedones, J.~Toledo, B.C.~van Rees and P.~Vieira, \emph{{The S-matrix bootstrap. Part I: QFT in AdS}}, \href{https://doi.org/10.1007/JHEP11(2017)133}{\emph{JHEP} {\bfseries 11} (2017) 133} [\href{https://arxiv.org/abs/1607.06109}{{\ttfamily 1607.06109}}].

\bibitem{Paulos:2017fhb}
M.F.~Paulos, J.~Penedones, J.~Toledo, B.C.~van Rees and P.~Vieira, \emph{{The S-matrix bootstrap. Part III: higher dimensional amplitudes}}, \href{https://doi.org/10.1007/JHEP12(2019)040}{\emph{JHEP} {\bfseries 12} (2019) 040} [\href{https://arxiv.org/abs/1708.06765}{{\ttfamily 1708.06765}}].

\bibitem{Hebbar:2020ukp}
A.~Hebbar, D.~Karateev and J.~Penedones, \emph{{Spinning S-matrix bootstrap in 4d}}, \href{https://doi.org/10.1007/JHEP01(2022)060}{\emph{JHEP} {\bfseries 01} (2022) 060} [\href{https://arxiv.org/abs/2011.11708}{{\ttfamily 2011.11708}}].

\bibitem{Guerrieri:2021tak}
A.~Guerrieri and A.~Sever, \emph{{Rigorous Bounds on the Analytic S Matrix}}, \href{https://doi.org/10.1103/PhysRevLett.127.251601}{\emph{Phys. Rev. Lett.} {\bfseries 127} (2021) 251601} [\href{https://arxiv.org/abs/2106.10257}{{\ttfamily 2106.10257}}].

\bibitem{Antunes:2023irg}
A.~Antunes, M.S.~Costa and J.~Pereira, \emph{{Exploring inelasticity in the S-matrix Bootstrap}}, \href{https://doi.org/10.1016/j.physletb.2023.138225}{\emph{Phys. Lett. B} {\bfseries 846} (2023) 138225} [\href{https://arxiv.org/abs/2301.13219}{{\ttfamily 2301.13219}}].

\bibitem{He:2023lyy}
Y.~He and M.~Kruczenski, \emph{{Bootstrapping gauge theories}}, \href{https://doi.org/10.1103/PhysRevLett.133.191601}{\emph{Phys. Rev. Lett.} {\bfseries 133} (2024) 191601} [\href{https://arxiv.org/abs/2309.12402}{{\ttfamily 2309.12402}}].

\bibitem{Haring:2023zwu}
K.~H{\"a}ring and A.~Zhiboedov, \emph{{The stringy S-matrix bootstrap: maximal spin and superpolynomial softness}}, \href{https://doi.org/10.1007/JHEP10(2024)075}{\emph{JHEP} {\bfseries 10} (2024) 075} [\href{https://arxiv.org/abs/2311.13631}{{\ttfamily 2311.13631}}].

\bibitem{Bhat:2024agd}
F.~Bhat, D.~Chowdhury, A.P.~Saha and A.~Sinha, \emph{{Bootstrapping string models with entanglement minimization and machine learning}}, \href{https://doi.org/10.1103/PhysRevD.111.066013}{\emph{Phys. Rev. D} {\bfseries 111} (2025) 066013} [\href{https://arxiv.org/abs/2409.18259}{{\ttfamily 2409.18259}}].

\bibitem{Eckner:2024ggx}
C.~Eckner, F.~Figueroa and P.~Tourkine, \emph{{Regge bootstrap: From linear to nonlinear trajectories}}, \href{https://doi.org/10.1103/PhysRevD.111.126005}{\emph{Phys. Rev. D} {\bfseries 111} (2025) 126005} [\href{https://arxiv.org/abs/2401.08736}{{\ttfamily 2401.08736}}].

\bibitem{Guerrieri:2024jkn}
A.~Guerrieri, K.~H{\"a}ring and N.~Su, \emph{{From data to the analytic S-matrix: A Bootstrap fit of the pion scattering amplitude}},  \href{https://arxiv.org/abs/2410.23333}{{\ttfamily 2410.23333}}.

\bibitem{He:2025gws}
Y.~He and M.~Kruczenski, \emph{{The Gauge Theory Bootstrap: Predicting pion dynamics from QCD}},  \href{https://arxiv.org/abs/2505.19332}{{\ttfamily 2505.19332}}.

\bibitem{deRham:2025vaq}
C.~de~Rham, A.J.~Tolley, Z.-H.~Wang and S.-Y.~Zhou, \emph{{Primal S-matrix bootstrap with dispersion relations}}, \href{https://doi.org/10.1007/JHEP01(2026)027}{\emph{JHEP} {\bfseries 01} (2026) 027} [\href{https://arxiv.org/abs/2506.22546}{{\ttfamily 2506.22546}}].

\bibitem{Guerrieri:2024ckc}
A.~Guerrieri, A.~Homrich and P.~Vieira, \emph{{Multiparticle Flux-Tube S-matrix Bootstrap}}, \href{https://doi.org/10.1103/PhysRevLett.134.041601}{\emph{Phys. Rev. Lett.} {\bfseries 134} (2025) 041601} [\href{https://arxiv.org/abs/2404.10812}{{\ttfamily 2404.10812}}].

\bibitem{Correia:2025uvc}
M.~Correia, A.~Georgoudis and A.L.~Guerrieri, \emph{{Cross-Section Bootstrap: Unveiling the Froissart Amplitude}},  \href{https://arxiv.org/abs/2506.04313}{{\ttfamily 2506.04313}}.

\bibitem{Chakraborty:2024ciu}
D.~Chakraborty, S.~Chattopadhyay and R.S.~Gupta, \emph{{Towards the HEFT-hedron: the complete set of positivity constraints at NLO}},  \href{https://arxiv.org/abs/2412.14155}{{\ttfamily 2412.14155}}.

\bibitem{Guerrieri:2021ivu}
A.~Guerrieri, J.~Penedones and P.~Vieira, \emph{{Where Is String Theory in the Space of Scattering Amplitudes?}}, \href{https://doi.org/10.1103/PhysRevLett.127.081601}{\emph{Phys. Rev. Lett.} {\bfseries 127} (2021) 081601} [\href{https://arxiv.org/abs/2102.02847}{{\ttfamily 2102.02847}}].

\bibitem{Caron-Huot:2021rmr}
S.~Caron-Huot, D.~Mazac, L.~Rastelli and D.~Simmons-Duffin, \emph{{Sharp boundaries for the swampland}}, \href{https://doi.org/10.1007/JHEP07(2021)110}{\emph{JHEP} {\bfseries 07} (2021) 110} [\href{https://arxiv.org/abs/2102.08951}{{\ttfamily 2102.08951}}].

\bibitem{Henriksson:2022oeu}
J.~Henriksson, B.~McPeak, F.~Russo and A.~Vichi, \emph{{Bounding violations of the weak gravity conjecture}}, \href{https://doi.org/10.1007/JHEP08(2022)184}{\emph{JHEP} {\bfseries 08} (2022) 184} [\href{https://arxiv.org/abs/2203.08164}{{\ttfamily 2203.08164}}].

\bibitem{Caron-Huot:2024lbf}
S.~Caron-Huot and Y.-Z.~Li, \emph{{Gravity and a universal cutoff for field theory}}, \href{https://doi.org/10.1007/JHEP02(2025)115}{\emph{JHEP} {\bfseries 02} (2025) 115} [\href{https://arxiv.org/abs/2408.06440}{{\ttfamily 2408.06440}}].

\bibitem{Boisvert:2026sfh}
M.~Boisvert, W.~Knop and L.~Rastelli, \emph{{Where is tree-level heterotic string theory?}},  \href{https://arxiv.org/abs/2606.09980}{{\ttfamily 2606.09980}}.

\bibitem{Albert:2024yap}
J.~Albert, W.~Knop and L.~Rastelli, \emph{{Where is tree-level string theory?}}, \href{https://doi.org/10.1007/JHEP02(2025)157}{\emph{JHEP} {\bfseries 02} (2025) 157} [\href{https://arxiv.org/abs/2406.12959}{{\ttfamily 2406.12959}}].

\bibitem{Albert:2022oes}
J.~Albert and L.~Rastelli, \emph{{Bootstrapping pions at large N}}, \href{https://doi.org/10.1007/JHEP08(2022)151}{\emph{JHEP} {\bfseries 08} (2022) 151} [\href{https://arxiv.org/abs/2203.11950}{{\ttfamily 2203.11950}}].

\bibitem{Albert:2023jtd}
J.~Albert and L.~Rastelli, \emph{{Bootstrapping pions at large N. Part II. Background gauge fields and the chiral anomaly}}, \href{https://doi.org/10.1007/JHEP09(2024)039}{\emph{JHEP} {\bfseries 09} (2024) 039} [\href{https://arxiv.org/abs/2307.01246}{{\ttfamily 2307.01246}}].

\bibitem{Albert:2023seb}
J.~Albert, J.~Henriksson, L.~Rastelli and A.~Vichi, \emph{{Bootstrapping mesons at large N: Regge trajectory from spin-two maximization}}, \href{https://doi.org/10.1007/JHEP09(2024)172}{\emph{JHEP} {\bfseries 09} (2024) 172} [\href{https://arxiv.org/abs/2312.15013}{{\ttfamily 2312.15013}}].

\bibitem{Albert:2026xyz}
J.~Albert, D.~Kosva and L.~Rastelli, \emph{{Bootstrapping Pion Form Factors at Large $N$}},  \href{https://arxiv.org/abs/2606.19420}{{\ttfamily 2606.19420}}.

\bibitem{deRham:2026lvc}
C.~de~Rham, A.J.~Tolley, Z.-H.~Wang and S.-Y.~Zhou, \emph{{Unitary Dual-Resonance S-matrices}},  \href{https://arxiv.org/abs/2607.24922}{{\ttfamily 2607.24922}}.

\bibitem{froissart1961asymptotic}
M.~Froissart, \emph{{Asymptotic behavior and subtractions in the Mandelstam representation}}, \href{https://doi.org/10.1103/PhysRev.123.1053}{\emph{Phys. Rev.} {\bfseries 123} (1961) 1053}.

\bibitem{jin1964connection}
Y.~Jin and A.~Martin, \emph{Connection between the asymptotic behavior and the sign of the discontinuity in one-dimensional dispersion relations}, {\emph{Physical Review} {\bfseries 135} (1964) B1369}.

\bibitem{Haring:2022cyf}
K.~H{\"a}ring and A.~Zhiboedov, \emph{{Gravitational Regge bounds}}, \href{https://doi.org/10.21468/SciPostPhys.16.1.034}{\emph{SciPost Phys.} {\bfseries 16} (2024) 034} [\href{https://arxiv.org/abs/2202.08280}{{\ttfamily 2202.08280}}].

\bibitem{Mandelstam1959analytic}
S.~Mandelstam, \emph{Analytic properties of transition amplitudes in perturbation theory}, \href{https://doi.org/10.1103/PhysRev.115.1741}{\emph{Phys. Rev.} {\bfseries 115} (1959) 1741}.

\bibitem{ChewMandelstam1960}
G.F.~Chew and S.~Mandelstam, \emph{Theory of low-energy pion-pion interactions}, \href{https://doi.org/10.1103/PhysRev.119.467}{\emph{Physical Review} {\bfseries 119} (1960) 467}.

\bibitem{Chandrasekaran:2018qmx}
V.~Chandrasekaran, G.N.~Remmen and A.~Shahbazi-Moghaddam, \emph{{Higher-Point Positivity}}, \href{https://doi.org/10.1007/JHEP11(2018)015}{\emph{JHEP} {\bfseries 11} (2018) 015} [\href{https://arxiv.org/abs/1804.03153}{{\ttfamily 1804.03153}}].

\bibitem{Cheung:2025nhw}
C.~Cheung and G.N.~Remmen, \emph{{Multipositivity bounds for scattering amplitudes}}, \href{https://doi.org/10.1103/wt4x-2149}{\emph{Phys. Rev. D} {\bfseries 112} (2025) 016017} [\href{https://arxiv.org/abs/2505.05553}{{\ttfamily 2505.05553}}].

\bibitem{Cheung:2026lpv}
C.~Cheung, J.~Jeong, P.~Ko, A.~Pomarol, G.N.~Remmen and F.~Sciotti, \emph{{Multipositivity Constrains the Chiral Lagrangian}},  \href{https://arxiv.org/abs/2605.21582}{{\ttfamily 2605.21582}}.

\bibitem{Jeong:2026xzk}
J.~Jeong, \emph{{Partial Waves for Multipositivity}},  \href{https://arxiv.org/abs/2608.02719}{{\ttfamily 2608.02719}}.

\end{thebibliography}\endgroup


\end{document}